\documentclass[a4paper,aps,preprintnumbers,showpacs,twocolumn,superscriptaddress,nofootinbib]{revtex4-1}
\usepackage{url}

\usepackage{graphicx}
\usepackage{dcolumn}
\usepackage{bm}
\usepackage{color}
\usepackage{amssymb}
\usepackage{amsmath}
\usepackage{times}
\usepackage{hyperref}

\usepackage{latexsym}
\usepackage{times}
\usepackage{stmaryrd}

\newcommand{\cf}{{cf.}~}
\newcommand{\ie}{{i.e.,}~}
\newcommand{\eg}{{e.g.,}~}

\newcommand{\msun}{\,M_{\odot}}

\begin{document}

\title[Nonsingular electrodynamics of a rotating and moving black
    hole]{Nonsingular electrodynamics of a rotating black hole moving
    in an asymptotically uniform magnetic test field}

\author{Viktoriya~S.~Morozova}
\affiliation{
Max-Planck-Institut f{\"u}r Gravitationsphysik, Albert Einstein
Institut, Am M{\"u}hlenberg 1, 14476 Potsdam, Germany
}
\affiliation{
California Institute of Technology, 1200 E California Blvd.,
Pasadena, CA 91125, USA
}

\author{Luciano Rezzolla}
\affiliation{
Institut f{\"u}r Theoretische Physik, Max-von-Laue-Str. 1, 60438
Frankfurt, Germany }
\affiliation{
Max-Planck-Institut f{\"u}r Gravitationsphysik, Albert Einstein
Institut, Am M{\"u}hlenberg 1, 14476 Potsdam, Germany
}

\author{Bobomurat~J.~Ahmedov}
\affiliation{
Institute of Nuclear Physics, Ulughbek, Tashkent 100214,
Uzbekistan
}
\affiliation{
Ulugh Beg Astronomical Institute, Astronomicheskaya 33, Tashkent
100052, Uzbekistan
}
\affiliation{The Abdus Salam International Centre for
Theoretical Physics, I-34151 Trieste, Italy
}

\date{\today}

\begin{abstract}
We extend the Wald solution to a black hole that is also moving at
constant velocity. More specifically, we derive analytic solutions for
the Maxwell equations for a rotating black hole moving at constant speed
in an asymptotically uniform magnetic test field. By adopting Kerr-Schild
coordinates we avoid singular behaviours at the horizon and obtain a
complete description of the charge and current distributions in terms of
the black-hole spin and velocity. Using this solution, we compute the energy
losses expected when charged particles are accelerated along the magnetic
field lines, improving previous estimates that had to cope with singular
electromagnetic fields on the horizon. When used to approximate the
emission from binary black holes in a uniform magnetic field, our
estimates match reasonably well those from numerical-relativity
calculations in the force-free approximation.
\end{abstract}

\maketitle

\section{Introduction}

In 1974 Wald~\cite{Wald:74bh} derived the exact solution of the vacuum
Maxwell equations in a Kerr spacetime for an asymptotically uniform
magnetic field aligned with the rotational axis of the black
hole. Although far away from the black hole this solution is described
by a uniform magnetic field and by an electric field which scales like
$1/r^2$, where $r$ is the distance from the black hole, near the
horizon the electric field can be large and comparable with the
magnetic one. Since then, the properties of black holes immersed in an
external magnetic field have been studied extensively and by several
authors, \eg~\cite{Aliev:2002, Dhurandhar:1984, Petterson:1975,
  Chitre:1975, Aliev:1993, Karas:2009, Bicak:2007,
  Karas:2012}. Although the vacuum Wald solution has an unscreened
parallel electric field which can trigger pair production and drive
poloidal currents, the injected charges eventually redistribute along
the field lines and screen the parallel electric field, forming a
force-free magnetosphere.

On a parallel track, Goldreich and Julian~\cite{Goldreich:1969} analyzed
in 1969 the vacuum solution for a rotating neutron star with a dipolar
magnetic field aligned with the rotational axis. They argued that the
rotationally-induced electric field was strong enough to pull charged
particles from the stellar surface and, thus fill the surrounding space
with plasma. Using the force-free approximation to describe the produced
magnetosphere, Michel argued that an electromagnetically-driven wind
would carry away rotational energy and angular momentum of the
star~\cite{Michel1973}. Finally, in 1977 Blandford and
Znajek~\cite{Blandford1977} realized that the similarity between the
vacuum solution for a Kerr black hole and the vacuum solution for a
rotating neutron star implied the possibility of generating an
electromagnetically driven wind from a rotating black hole, provided that
space around the black hole could be filled with plasma. The literature
that has developed around this suggestion is too large to be reported
here and a useful list of references can be found
in~\cite{Komissarov:2009dn}.

More recently, attention has been paid to the possibility that a
significant electromagnetic signal can be produced during the inspiral
and merger of a binary of supermassive black holes. This scenario has
been considered both for spacetimes in
electrovacuum~\cite{Palenzuela:2009yr, Palenzuela:2009hx, Moesta:2009}
and for spacetimes filled by a tenuous plasma in the force-free
approximation~\cite{Palenzuela:2010a, Palenzuela:2010b,
  Neilsen:2010ax}. In particular, it was pointed out
in~\cite{Palenzuela:2010a} that during the inspiral in a force-free
plasma, a \textit{dual-jet} structure forms, as a generalization of the
Blandford-Znajek process to orbiting black holes, where the
electromagnetic energy flux is concentrated along the magnetic-field
lines. Under these conditions, the electromagnetic energy can accelerate
electrons and lead to synchrotron radiation. This picture has been
further refined in~\cite{Moesta2011,Alic:2012}, where it was shown that a
dual-jet structure is present but energetically subdominant with respect
to a non-collimated and predominantly quadrupolar emission, which is
similar to the one computed when the binary is in electrovacuum. A conclusive 
answer on whether dual jets can be launched and be used to detect
electromagnetic counterparts of the merger of supermassive black holes
will require further study and, in particular, a better understanding of
the interaction of a rotating black hole as it moves at relativistic
speeds in a magnetized and tenuous plasma. This is one of the main
motivations behind this work.

More specifically, we derive the analytic solution for the
electromagnetic fields in the vicinity of a moving and rotating black
hole at first order in the spin. In this respect, we extend not only the
Wald solution to the case of moving black holes, but also the recent
analysis of Lyutikov~\cite{Lyutikov:2011}, who has considered the
electrodynamics of a moving Schwarzschild black hole. Differently from
Wald and Lyutikov, however, we adopt a Kerr-Schild metric, thus obtaining
solutions which are regular everywhere except at the physical
singularity. Using this solution we compute the electromagnetic losses
expected when charged particles are accelerated along the magnetic field
lines and discuss how these depend on the black hole's spin and linear
velocities. Considering a reference supermassive black hole of mass
$M=10^8\,\msun$, immersed in an external asymptotically uniform magnetic
field of the order of $10^4\,\mathrm{G}$, we estimate an electromagnetic
luminosity $L_{_{\rm EM}} \sim 10^{44}\,\mathrm{erg\ s^{-1}}$.

The paper is organized as follows. In Sect.~\ref{SCHW} we first consider
the electromagnetic fields in the vicinity of nonrotating but moving
black hole in Kerr-Schild coordinates. Section \ref{KERR} generalizes the
exact Wald solution for the Kerr black hole to the case of Kerr-Schild
coordinates, while Section \ref{BoostedKERR} extends the analysis to the
most generic case of a moving and rotating black hole. In
Sect.~\ref{LOSS} we calculate the energy losses due to currents flowing
in the vicinity of the black hole's horizon and discuss the detectability
of this emission in the case of a supermassive black hole. Finally, a
summary of the results and our conclusions are presented in
Sect.~\ref{CONC}.

We use a spacetime signature $(-,+,+,+)$, with Greek indices running from
$0$ to $3$ and the Latin indices from $1$ to $3$. We also employ the
standard convention for the summation over repeated indices. Finally, all
the quantities are expressed in a system of units in which $c=G=1$,
unless otherwise stated.

\section{Moving Schwarzschild black hole in a uniform magnetic field}
\label{SCHW}

In order to obtain solutions of the Maxwell equations which are not
singular at the horizon, we write the black-hole solution in
Kerr-Schild coordinates expressing the line element
\begin{equation}
ds^2 = g_{\mu\nu} dx^{\mu} dx^{\nu}\,,
\end{equation}
in terms of a metric tensor whose nonzero components in spherical
polar coordinates are
\begin{subequations}
\label{metric}
\begin{align}
g_{tt} &= -\left(1-\frac{2Mr}{\Sigma}\right)\,, &
g_{tr} &=\frac{2Mr}{\Sigma}\,, \\
g_{t\phi} &= -\frac{2Mar\sin^2\theta}{\Sigma}\,, &
g_{rr} &= 1+\frac{2Mr}{\Sigma}\,, \\
g_{\theta\theta} &= \Sigma\,, &
g_{r \phi} &= -a\sin^2\theta\left(1+\frac{2Mr}{\Sigma}\right)\,, \\
g_{\phi\phi} &= \frac{A\sin^2\theta}{\Sigma}\,,&
\end{align}
\end{subequations}
and where
\begin{subequations}
\label{defs}
\begin{align}
& \Sigma\equiv r^2+a^2\cos^2\theta\,,\quad \Delta \equiv r^2-2Mr+a^2\,,
\\
& A \equiv (r^2+a^2)^2-a^2\Delta \sin^2\theta\,,
\\
& a  \equiv J/M\,,
\end{align}
\end{subequations}
$J$ being the angular momentum of the black hole with mass $M$.

As customary when dealing with rotating black holes, we consider zero
angular momentum observers (ZAMOs) as those with a future-directed
unit vector orthogonal to a $t=\mathrm{const}$ hypersurface. The
components of the four-velocity for such observers in the Kerr-Schild
coordinates at second order in the spin, \ie at ${\cal O}(a^2)$, are
then given by~\cite{Takahashi:2007, Takahashi:2008}
\begin{subequations}
\begin{align}
\label{ZAMO_cova} 
&(u_{_{\textrm{ZAMO}}})_{\alpha}
=\left\{-\frac{1}{\tilde{N}}\left(1+
\frac{Ma^2\cos^2\theta}{\tilde{N}^2r^3}\right),0,0,0\right\}\
, \\
\label{ZAMO_contra}
&(u_{_{\textrm{ZAMO}}})^{\alpha}=
\Bigg\{\tilde{N}\left(1-\frac{Ma^2\cos^2\theta}{r^3\tilde{N}^2}\right),\nonumber
\\ 
& \hskip 2.0cm
-\frac{2M}{r\tilde{N}}\left(1-
\frac{a^2\cos^2\theta(r+M)}{r^3\tilde{N}^2}\right),0,0\Bigg\}\,,
\end{align}
\end{subequations}
where we have introduced the functions $\tilde{N}^2 \equiv 1+{2M}/{r}$
and ${N}^2 \equiv 1-{2M}/{r}$. A well-known property of ZAMO observers
is that, despite having zero angular momentum and thus freely falling
towards the black hole, they also rotate around it, dragged by the
general rotation of the spacetime.

To simplify the analysis and perform a more direct comparison with the
results of Lyutikov~\cite{Lyutikov:2011}, we will consider in this
Section the electromagnetic fields in the vicinity of a
\textit{nonrotating} black hole (\ie with $a=0$) as it is moving at
constant velocity in an asymptotically uniform magnetic field with
velocity $\beta$. We note that a similar analysis in Kerr-Schild
coordinates was already presented in the Appendix of
Ref.~\cite{Lyutikov:2011}, where, however, the $\phi$-components of the
electric and magnetic fields were neglected, thus leading to a result
which is effectively valid only at first order in the velocity.

Following Lyutikov~\cite{Lyutikov:2011}, we assume the black hole to be
immersed in external magnetic field $\boldsymbol{\vec{B}}$ with value
$B_0$, which is asymptotically uniform and directed along the
$z$-direction in a Cartesian coordinate system, \ie $B^i = B_0(0, 0, 1)$
for $x^i \to \infty$. We also assume that the black hole is moving at
constant speed orthogonally to the magnetic field with velocity $v=\beta$
in the negative $y$-direction, \ie $v^i = \beta(0, -1,
0)$\footnote{In practice it is more convenient to use a reference
    frame comoving with the black hole and hence consider the magnetic field as
    moving with respect to the black hole. Of course Lorentz invariance
    guarantees that the two descriptions are equivalent.}. An observer
comoving with the black hole will measure an electric field
$\boldsymbol{\vec{E}} = -\boldsymbol{\vec{v}} \times
\boldsymbol{\vec{B}}$ which is asymptotically uniform and along the
$x$-direction (hence orthogonal to the magnetic field) and with magnitude
\begin{equation}
\label{eq:E0}
E_0=\beta B_0 \,,
\end{equation}
at spatial infinity. Near the black hole, however, the electromagnetic
field will be influenced by the spacetime curvature and its form can be
found via the solution of the vacuum Maxwell equations
\begin{subequations}
\begin{align}
\label{Max_1}
&\partial_{\nu}\left(\sqrt{-g}\ ^*\!F^{\mu\nu}\right)=0\,, \\
\label{Max_2}
&\partial_{\nu}\left(\sqrt{-g}F^{\mu\nu}\right)=0\,,
\end{align}
\end{subequations}
where $g$ is the determinant of the metric tensor, $F_{\alpha\beta}$ is the Faraday tensor and can be expressed in
terms of the electromagnetic vector potential $A_{\alpha}$ as
$F_{\alpha\beta} \equiv A_{\beta\,,\alpha}-A_{\alpha\,,\beta}$, while the asterisk denotes 
the dual of the Faraday tensor. Using the
asymptotic values of the electromagnetic fields $E_0$ and $B_0$, the
covariant components of the vector potential in Kerr-Schild coordinates
are (see Appendix~\ref{Ap1} for details)
\begin{eqnarray}
\label{aLut}
&&A_t=N^2E_0r\sin\theta\cos\phi\,,\\
\label{aLuphi}
&&A_{\phi}=\frac{B_0}{2}r^2\sin^2\theta\,,\\
\label{aLur}
&&A_r=-2ME_0\sin\theta\cos\phi\,.
\end{eqnarray}
From these components, it is then possible to compute the Faraday
tensor and thus the coordinate components of the electromagnetic fields
\begin{equation}
E_{\alpha} \equiv F_{\alpha\beta}u^{\beta}\,,\qquad
B^{\alpha} \equiv 
\frac{1}{2}\eta^{\alpha\beta\gamma\delta}u_{\beta}F_{\gamma\delta}\,,
\end{equation}
where $\eta^{\alpha\beta\gamma\delta}=
-({1}/{\sqrt{-g}})\epsilon_{\alpha\beta\gamma\delta}$ and $\epsilon_{\alpha\beta\gamma\delta}$
is the Levi-Civita symbol. The corresponding physical components of
electromagnetic fields can be found after projecting the electromagnetic
field components on the tetrads carried out by the ZAMOs, \ie through the
relations $E_{\hat{\alpha}} = e_{~\hat{\alpha}}^{\beta}E_{\beta}$, and
$B^{\hat{\alpha}} = e^{\hat{\alpha}}_{~\beta}B^{\beta}$, where
$\boldsymbol{e}^{\hat{\beta}}$ is the ZAMO tetrad in the slow-rotation
approximation [see expressions~\eqref{tetrads1} in Appendix~\ref{Ap3}
  and Ref.~\cite{Takahashi:2007, Takahashi:2008} for the full
  expressions]. A little of algebra then yields
\begin{eqnarray}
\label{Ehatr}
&&E_{\hat{r}}=E_0\sin\theta\cos\phi\,,\\
\label{Ehattheta}
&&E_{\hat{\theta}}=\frac{1}{\tilde{N}}E_0\cos\theta\cos\phi\ \,, \\
\label{Ehatphi}
&&E_{\hat{\phi}}=\frac{1}{\tilde{N}}
\left(-E_0\sin\phi+\frac{2M}{r}B_0\sin\theta\right)\,,
\end{eqnarray}
and
\begin{eqnarray}
\label{Bhatr}
&&B_{\hat{r}}=-B_0\cos\theta\,,  \\
\label{Bhattheta}
&&B_{\hat{\theta}}=\frac{1}{\tilde{N}}\left(B_0\sin\theta-
\frac{2M}{r}E_0\sin\phi\right)\,, \\
\label{Bhatphi}
&&B_{\hat{\phi}}=-\frac{2M}{r\tilde{N}}E_0\cos\theta\cos\phi\,.
\end{eqnarray}

\begin{figure}
\includegraphics[width=0.45\textwidth]{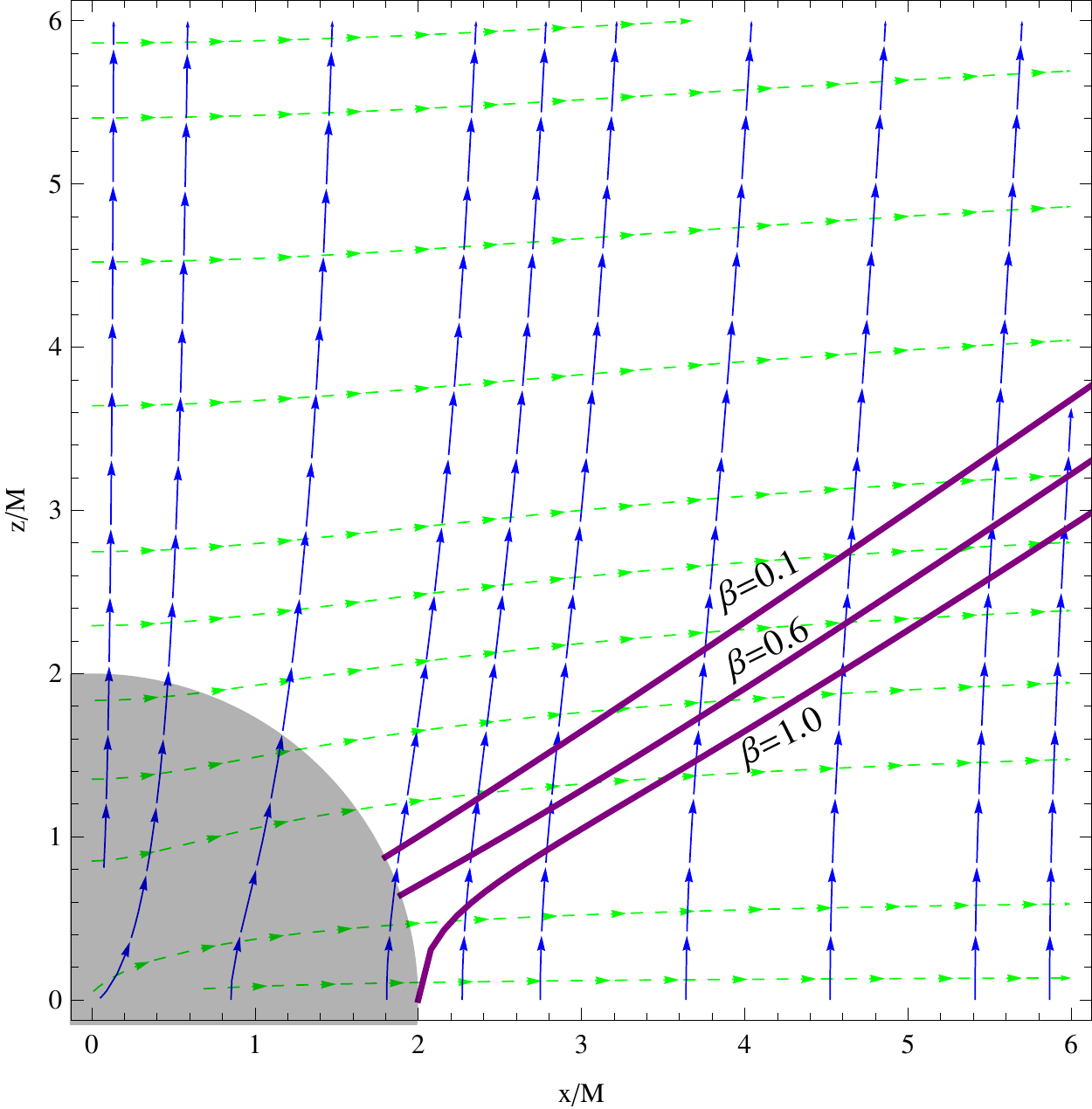}
\caption{Electromagnetic fields in the vicinity of a Schwarzschild black
  hole moving in the negative $y$-direction. Shown with blue and green
  lines are field lines of the magnetic and electric field, respectively,
  while the horizon is represented as a gray-shaded area. Also shown with
  thick purple lines are the charge separatrixes $\rho_{\rm ind}=0$ for
  the different values of the parameter $\beta$ in the plane
  $\phi=0$. The highest separatrix corresponds to the case $\beta=0.1$,
  the middle one to the case $\beta=0.6$ and the lowest line to the case
  $\beta=1$ and these move toward the equatorial plane for growing
  $\beta$.} \label{Fig1}
\end{figure}

Note that in contrast with what happens for Boyer-Linquist coordinates,
the electric and magnetic fields have a nonzero covariant
$\phi$-component in Kerr-Schild coordinates. Equation (\ref{Ehatphi}), in
particular, shows that the $\phi$-component of the electric field
contains a term which is proportional to $B_0$ but not to $\beta$ and
which therefore does not vanish when $\beta=0$. This is related to the
fact that ZAMO observers also have a radial motion, which is responsible
for the appearance of this electric field even if the black hole is not
moving; clearly this term vanishes at infinity. An example of the
electromagnetic field structure is shown in Fig.~\ref{Fig1}, where the
(green) solid and (blue) dashed vector field lines refer to the electric
and magnetic fields, respectively, while the horizon is represented as a
gray-shaded area. Note that the electric and magnetic fields are
orthogonal only asymptotically and thus that a component of the electric 
field $\vec{\boldsymbol{E}}_{\parallel}$ parallel to the magnetic field is 
present near the horizon, which can lead to particle acceleration there (see discussion
below). 

Also shown with thick purple lines are the charge separatrixes $\rho_{\rm
  ind}=0$ for the different values of the parameter $\beta$ in the plane
$\phi=0$. The highest separatrix corresponds to the case $\beta=0.1$, the
middle one to the case $\beta=0.6$ and the lowest line to the case
$\beta=1$ and these move toward the equatorial plane for growing
$\beta$. This figure should be compared with the corresponding Fig.~1
of~\cite{Lyutikov:2011}, where the electromagnetic fields are however
divergent at the horizon and are not reported inside the black
hole\footnote{\label{footnote:dual}As remarked in~\cite{Lyutikov:2011},
  the electric and magnetic fields described by expressions
  (\ref{Ehatr})--(\ref{Bhatphi}) have the same form after a dual
  transformation. As a result, a dual solution exists describing a
  Schwarzschild black hole moving in the $-y$ direction in an
  asymptotically uniform magnetic field directed along the $-x$
  direction; in this case, the electric field $E_0=\beta B_0$ will be
  directed along the $z$ direction (see Appendix~\ref{ApD} for details).}

Using expressions~\eqref{Ehatr}--\eqref{Bhatphi} we can compute the
electromagnetic-field \textit{invariants}~\cite{Wald:74bh}
\begin{equation}
B^2-E^2=\frac{1}{2}F_{\mu\nu}F^{\mu\nu}\,, \qquad
\boldsymbol{\vec{E}}\cdot\boldsymbol{\vec{B}}
=\frac{1}{4} \ ^*\!F_{\mu\nu}F^{\mu\nu}\,.
\end{equation}
It is then
straightforward to define the electric-field component parallel to
the magnetic field as
\begin{equation}
\boldsymbol{\vec{E}}_{\parallel} \equiv
\frac{\boldsymbol{\vec{E}}\cdot
\boldsymbol{\vec{B}}}{B^2}\boldsymbol{\vec{B}}\,,
\end{equation}
and thus the induced charge density necessary to screen this electric
field component is
\begin{equation}
-\rho_{\rm ind} \equiv \frac{1}{4\pi}\nabla\cdot \boldsymbol{\vec{E}}_{\parallel}\,.
\end{equation}

The explicit expressions for the invariants $B^2-E^2$ and
$\boldsymbol{\vec{E}}\cdot\boldsymbol{\vec{B}}$, for the parallel
electric field $\boldsymbol{\vec{E}}_{\parallel}$ and for the induced
charge density $\rho_{\rm ind}$ in Kerr-Schild coordinates have already
been reported by Lyutikov~\cite{Lyutikov:2011}, which however are
incomplete since they do not include the $\phi$-components of the
electric and magnetic fields. The corresponding full expressions are
presented in Appendix~\ref{Ap2} [\cf Eq.~\eqref{rho_full}], while we
report below the angular distribution of the induced charge density at
first order in the velocity $\beta$.
\begin{widetext}
\begin{eqnarray}
\label{rho_ind}
&&\rho_{\rm ind}(r,\theta) = \frac{E_0M\sin\theta\cos\phi}{\pi
r^2\tilde{N}\left(1+{2M\cos^2\theta}/{r}\right)^2}
\Bigg(1-\frac{3}{2}\sin^2\theta+\frac{4M}{r}
-\frac{5M}{r}\sin^2\theta+
\frac{4M^2}{r^2}\cos^4\theta+\frac{2M^2}{r^2}\cos^2\theta\sin^2\theta\Bigg)\,.
\end{eqnarray}
\end{widetext}

Of particular interest in the function~\eqref{rho_ind} is the
``separatrix'', namely: the line where the function goes to zero and
hence where the charge distribution changes sign. The location of such
separatrix is independent of the velocity only when the induced charge
distribution is taken at first order in $\beta$ [\cf
  Eq.~\eqref{rho_ind}]; however, if the complete expression is
considered, then the separatrix is a function of the velocity and moves
towards the equatorial plane with increasing $\beta$ [\cf
  Eq.~\eqref{rho_full}]. This is shown with the thick solid lines in
Fig.~\ref{Fig1}, which refer to velocities $\beta=0.1, 0.6, 1$,
respectively. This effect was not remarked in~\cite{Lyutikov:2011}, where
only the first-order expression for the induced charge distribution was
considered. Also shown in Fig.~\ref{Fig1} are the magnetic-field lines
(blue lines) and the electric-field lines (green lines), which tend to be
orthogonal at sufficient distances from the black hole horizon.

The following Section will be dedicated to extending the analysis carried
out here to the case of a spinning black hole in the Kerr-Schild
coordinates.

\section{Wald solution in Kerr-Schild coordinates}
\label{KERR}

As a warm-up exercise, we re-derive the Wald solution~\cite{Wald:74bh} in
Kerr-Schild coordinates and thus for a stationary black hole, as this
will be one of the building blocks to obtain the full solution. Following
Wald~\cite{Wald:74bh}, we exploit the existence in the spacetime
(\ref{metric}) of a timelike Killing vector $\xi^\alpha_{(t)}$ and of a
spacelike one $\xi^\alpha_{(\phi)}$, which reflect the stationarity and
the axial symmetry of the black hole solution and such that they {satisfy
  the} Killing equation
\begin{equation}
\label{ke} 
\xi_{\alpha ;\beta}+\xi_{\beta;\alpha}=0\,.
\end{equation}
Following again Wald~\cite{Wald:74bh}, we can use Eq.~\eqref{ke} and
the tensor identity
\begin{equation}
\xi_{\alpha;\beta;\gamma}-\xi_{\alpha;\gamma;\beta} =-\xi^{\delta} R_{\delta\alpha\beta\gamma} \,,
\end{equation}
where $R_{\delta\alpha\beta\gamma}$ is the Riemann curvature tensor,
to obtain
\begin{equation}
\label{boxxi}
\Box \xi^\alpha = \xi^{\alpha ;\beta}_{\ \ \ \ \ ;\beta}=R^{\alpha}_{\ \delta}\xi^{\delta}\,,
\end{equation}
where we have permuted cyclically the indices and added the resulting
equations after exploiting the symmetries of the Riemann
tensor. Because the right-hand side of Eq.~\eqref{boxxi} vanishes in
vacuum, we obtain a wave equation for the Killing vector
$\xi^{\alpha}$. Since a Killing vector generates a solution of the
Maxwell equations, we set the vector potential to be a linear
combination of the Killing vectors
\begin{equation}
\label{pots} 
A^\alpha=C_2 \xi^\alpha_{(t)}+C_3
\xi^\alpha_{(\phi)}\,,
\end{equation}
and it is then easy to show that~\eqref{pots} is a solution of the
Maxwell equations $F^{\mu\nu}_{\ ;\nu}=0$ and that $\Box A^\mu=0$ in
the Lorentz gauge $A^{\mu}_{\ ;\mu} = 0$~\cite{Wald:74bh}.

Using the ansatz~(\ref{pots}), the integration constant $C_3$ is found
from the requirement that the black hole is immersed into an
asymptotically uniform magnetic field, which is parallel to its axis of
rotation. This procedure was adopted by Wald in~\cite{Wald:74bh} using
Boyer-Lindquist coordinates and of course it can be applied also in the
case of Kerr-Schild coordinates. More specifically, once the
$z$-component of the magnetic field is evaluated in the orthonormal basis
of a ZAMO observer at spatial infinity, it is sufficient to require that
this value is equal to $B_0$ to obtain the integration constant as
$C_3=B_0/2$.

The remaining integration constant can instead be found after imposing
charge neutrality as evaluated through a surface integral across a
spherical surface at spatial infinity, \ie
\begin{align}
\label{C2const}
4\pi Q=0 &= \frac{1}{2}\oint
F^{\alpha\beta}{_*dS}_{\alpha\beta} \nonumber \\ 
 &  =C_2 \oint
\xi_{(t)}^{\alpha;\beta}{_*dS}_{\alpha\beta} + 
\frac{B}{2}\oint\xi_{(\phi)}^{\alpha;\beta}{_*dS}_{\alpha\beta}\,,
\end{align}
where $dS_{\alpha \beta}$ is the infinitesimal element on the
2-sphere. Fortunately, the two integrals in~(\ref{C2const}) have
simple analytic solutions, namely~\cite{Poisson04a}:
\begin{align}
&\oint \xi_{(t)}^{\alpha;\beta}{_*dS}_{\alpha\beta} = -8\pi M\,, &
&\oint\xi_{(\phi)}^{\alpha;\beta}{_*dS}_{\alpha\beta} = 16 \pi J\,,
\end{align}
so that it is then straightforward to derive that $C_2=aB_0$. 

With the integration constants known, the explicit expressions for the
the covariant component of the 4-vector potential $A_\alpha$ will take
the form
\begin{eqnarray}
\label{potential0}
A_t&=&-aB_0\left(1-\frac{Mr}{\Sigma}(1+\cos^2\theta)\right)\,,\\
\label{potential1}
A_r&=&\frac{aB_0}{2}\left(\frac{2Mr}{\Sigma}(1+\cos^2\theta)-
\sin^2\theta\right)\,,\\
A_{\phi}&=&\frac{B_0\sin^2\theta}{2\Sigma}\left(A-4Ma^2r\right)\,.
\end{eqnarray}

Using equations~(\ref{pots}) and \eqref{potential0} it is also possible
to compute what is the maximum charge that the black hole can support,
following the arguments already given in~\cite{Wald:74bh}. In particular,
it is sufficient to recall that the electrostatic energy of a particle
with charge $q_e$ in the vicinity of the black hole is simply given by
$q_e A^{\mu}\xi_{\mu\ (t)}$. Because the Killing vector $\xi_{\mu\ (t)}$
becomes spacelike inside the ergoregion defined by $g_{tt}=0$, it is more
convenient to introduce a new Killing vector which is regular at the
horizon~\cite{Carter71,Aliev:1993}
\begin{equation}
(\psi_{\mu})_{\rm h}=\xi_{\mu(t)}+\Omega_{\rm h}\xi_{\mu(\phi)}\,,
\end{equation}
where
\begin{equation}
(\Omega)_{\rm h} \equiv \frac{a}{2Mr_+}\,,\quad 
r_+ \equiv M+\sqrt{M^2-a^2}\,.
\end{equation}
The change in the electrostatic energy of the charged particle as it falls
from infinity down to the black-hole horizon is thus given by
\begin{align}
\label{eq:ese}
\epsilon &= q_e\left\{ (A^\mu)_{\rm h} \left[(\xi_{\mu(t)})_{\rm h}
+(\Omega)_{\rm h}(\xi_{\mu(\phi)})_{\rm h} \right]
- (A^\mu)_{\rm \infty}(\xi_{\mu(t)})_{\infty}\right\} \nonumber \\
&=- q_e a B_0\,,
\end{align}
where the indices ${\rm h}$ and $\infty$ refer to quantities evaluated at
the horizon and at infinity, respectively. Assuming now that the black
hole is not electrically neutral, but instead possesses the net electric
charge $Q$, Eq.~\eqref{eq:ese} would have to be modified to add for the
additional electrostatic contribution induced by the black hole's charge,
\ie
\begin{equation}
\label{eq:ese_c}
\epsilon = \frac{q_e}{2M}\left(Q - 2 M a B_0\right)\,.
\end{equation}
At equilibrium, $\epsilon = 0$, so that it is possible to deduce the
upper limit for test electric charge as accreted by the rotating black hole
as~\cite{Wald:74bh}
\begin{equation}
Q=2aMB_0\,.
\end{equation}
Clearly, this charge vanishes for a nonrotating black hole.

Collecting things together we deduce that the nonvanishing components
of the Faraday tensor are then given by
\begin{widetext}
\begin{eqnarray}
F_{rt}&=&-\frac{aMB_0}{\Sigma^2}(r^2-a^2\cos^2\theta)(1+\cos^2\theta)\,,\\
F_{\theta t}&=&-\frac{2aMrB_0}{\Sigma^2}(r^2-a^2)\sin\theta\cos\theta\,,  \\
F_{\theta r}&=&-aB_0\left[\frac{2Mr}{\Sigma^2}(r^2-a^2)+1\right]\sin\theta\cos\theta\,, \\
F_{r \phi}&=&\frac{B_0}{\Sigma^2}\Bigg[2\Sigma r^3-Ar+4Ma^2r^2 + a^2\Sigma(r-M)(1+\cos^2\theta)\Bigg]\sin^2\theta\,, \\
F_{\theta \phi}&=&\frac{B_0}{\Sigma^2}\Bigg[(r^2+a^2)(A-4Ma^2r)
-a^2\Delta\Sigma\sin^2\theta\Bigg]\sin\theta\cos\theta\,. 
\end{eqnarray}

At this point it is straightforward to compute the expressions for the
orthonormal components of the electromagnetic fields measured by the ZAMO
observers. Using the four-velocity
components~\eqref{ZAMO_cova}--\eqref{ZAMO_contra}, which are at first
order in the spin parameter, the electric field will be given by
\begin{eqnarray}
\label{e1} && E^{\hat r}
=-\frac{aMB_0}{\sqrt{A}\Sigma^3}\Bigg\{
\frac{A(r^2-a^2\cos^2\theta)(1+\cos^2\theta)}{\left(1+2Mr/\Sigma\right)} 
-\Bigg[2\Sigma r^4-Ar^2+4Ma^2r^3 + 
a^2r\Sigma(r-M)(1+\cos^2\theta)\Bigg]\sin^2\theta\Bigg\}\,,\\
\label{e2} && E^{\hat
\theta}=\frac{2a^3MrB_0}{\Sigma^{2}\left(\Sigma+2Mr\right)^{1/2}}(1+\cos^2\theta)\sin\theta\cos\theta
 \,,\\
\label{e3} &&
E^{\hat\phi}=\frac{2MrB_0\sin\theta}{\Sigma^2\sqrt{A}\left(\Sigma+2Mr\right)^{1/2}}\Bigg[2\Sigma
r^3-Ar+4Ma^2r^2 + a^2\Sigma(r-M)(1+\cos^2\theta)\Bigg]\,,
\end{eqnarray}
while the magnetic field components will be given by 
\begin{eqnarray}
\label{m1} && B^{\hat r} =\frac{2B_0\cos\theta}{\Sigma
\sqrt{A}\Xi}\Bigg[(r^2+a^2)(A-4Ma^2r)-a^2\Delta\Sigma\sin^2\theta\Bigg]
\,,
\\
\label{m2} && B^{\hat\theta} =
-\frac{2B_0}{\Sigma\left(\Sigma+2Mr\right)^{1/2}
\Xi} \Bigg[2\Sigma r^3-Ar+4Ma^2r^2 +
a^2\Sigma(r-M)(1+\cos^2\theta)\Bigg]\sin\theta\,, \\
\label{m3} && B^{\hat\phi} =  \frac{2
aB_0\sin\theta\cos\theta}{\sqrt{A}\Sigma^2\left(\Sigma+2Mr\right)^{1/2}\Xi}
\Bigg\{{2MrA}(r^2-a^2) 
+A-{\left(\Sigma+2Mr\right)}
\left[(r^2+a^2)(A-4Ma^2r)-a^2\Delta\Sigma\sin^2\theta\right]\Bigg\}\,.
\end{eqnarray}
\end{widetext}
where $\Xi \equiv (a^2+2r^2+a^2\cos2\theta)$. It is then easy to
recognize that the geometric structure of the electric field is complex
and with a non trivial dependence from the black-hole's angular momentum
$a$. The radial and polar components of the electric field are due to the
dragging of inertial frames and the $\phi$-component is due to the radial
fall of ZAMO observer in the Kerr-Schild coordinates. In the limit of a
flat spacetime, \ie for \hbox{$M/r\rightarrow 0$, and $Ma/r^2\rightarrow
  0$}, expressions (\ref{e1})--(\ref{m2}) give
\begin{eqnarray}
\label{limit_B_1} 
&& \lim_{M/r, Ma/r^2\rightarrow 0} B^{\hat r}=-B_0\cos\theta \,,  
 \lim_{M/r, Ma/r^2\rightarrow 0}
B^{\hat\theta}=B_0\sin\theta   \,, \nonumber
\\ \nonumber \\
\label{limit_E} 
&& \lim_{M/r, Ma/r^2\rightarrow 0}B^{\hat\phi}=0\,,
\qquad \lim_{M/r, Ma/r^2\rightarrow 0} E^{\hat j}=0\,. 
\end{eqnarray}
which coincide with the solutions for the homogeneous magnetic field in a
Minkowski spacetime. Also, we have checked that the expression
(\ref{e1})--(\ref{m2}) satisfy the Maxwell
equations~\eqref{Max_1}--\eqref{Max_2}.

\section{Moving Wald solution in Kerr-Schild coordinates}
\label{BoostedKERR}

We next consider the solution of the Maxwell equations for a moving and
spinning black hole, thus extending the solutions of Wald and Lyutikov to
the most generic case. For simplicity, we will consider the black-hole
spin axis to be along the $z$-direction (the same direction of the
asymptotically uniform magnetic field) and its velocity to be instead in
the negative $y$-direction; more generic configurations can be easily
derived but would only introduce additional trigonometric corrections. In
addition, to keep the expressions in a compact form, we will consider the
black-hole metric in the slow-rotation approximation, that is, at
first-order in the spin parameter $a$. In this case, the metric
functions~\eqref{metric} reduce to
\begin{subequations}
\label{metric_sr}
\begin{align}
g_{tt} &= -N^2\,, &
g_{tr} &= g_{rt}=\frac{2M}{r}\,, \\ 
g_{t \phi} &= g_{\phi t}=-\frac{2Ma\sin^2\theta}{r}\,, &
g_{rr}    &=1+\frac{2M}{r}\,, & \!\!\!g_{\theta\theta} &= r^2\,, \\ 
g_{r \phi} &= g_{\phi r}=-a\tilde{N}^{2}\sin^2\theta\,, &
g_{\phi\phi} &= r^2\sin^2\theta\,. 
\end{align}
\end{subequations}

To obtain the moving Wald solution in Kerr-Schild coordinates, we exploit
the linearity of electromagnetic equations and write the solution of
Maxwell equations in the spacetime~(\ref{metric}) as the superposition of
the Wald solution discussed above with the one relative to a moving black
hole (\ref{aLut})--(\ref{aLur}), so that the covariant components of the
vector potential can be written as
\begin{subequations}
\label{eq:vecpot_boostedkerr}
\begin{align}
\label{A_t} 
A_t &= -aB_0\left(N^2+\frac{M\sin^2\theta}{r}\right) +N^2E_0r\sin\theta\cos\phi \,,\\
\label{A_r}
A_r &= aB_0\left(\frac{2M}{r}-\frac{\tilde{N}^2\sin^2\theta}{2}\right) - 2ME_0\sin\theta\cos\phi \,, \\
\label{A_phi} 
A_{\phi} &= \frac{B_0}{2}r^2\sin^2\theta \,,
\end{align}
\end{subequations}
where we recall that $E_0=\beta B_0$ is the electric field
  measured by an observer comoving with the black hole and the axial
symmetry is clearly broken through the new dependence on the azimuthal angle
$\phi$.

\begin{figure*}
\begin{center}
\includegraphics[width=0.235\textwidth]{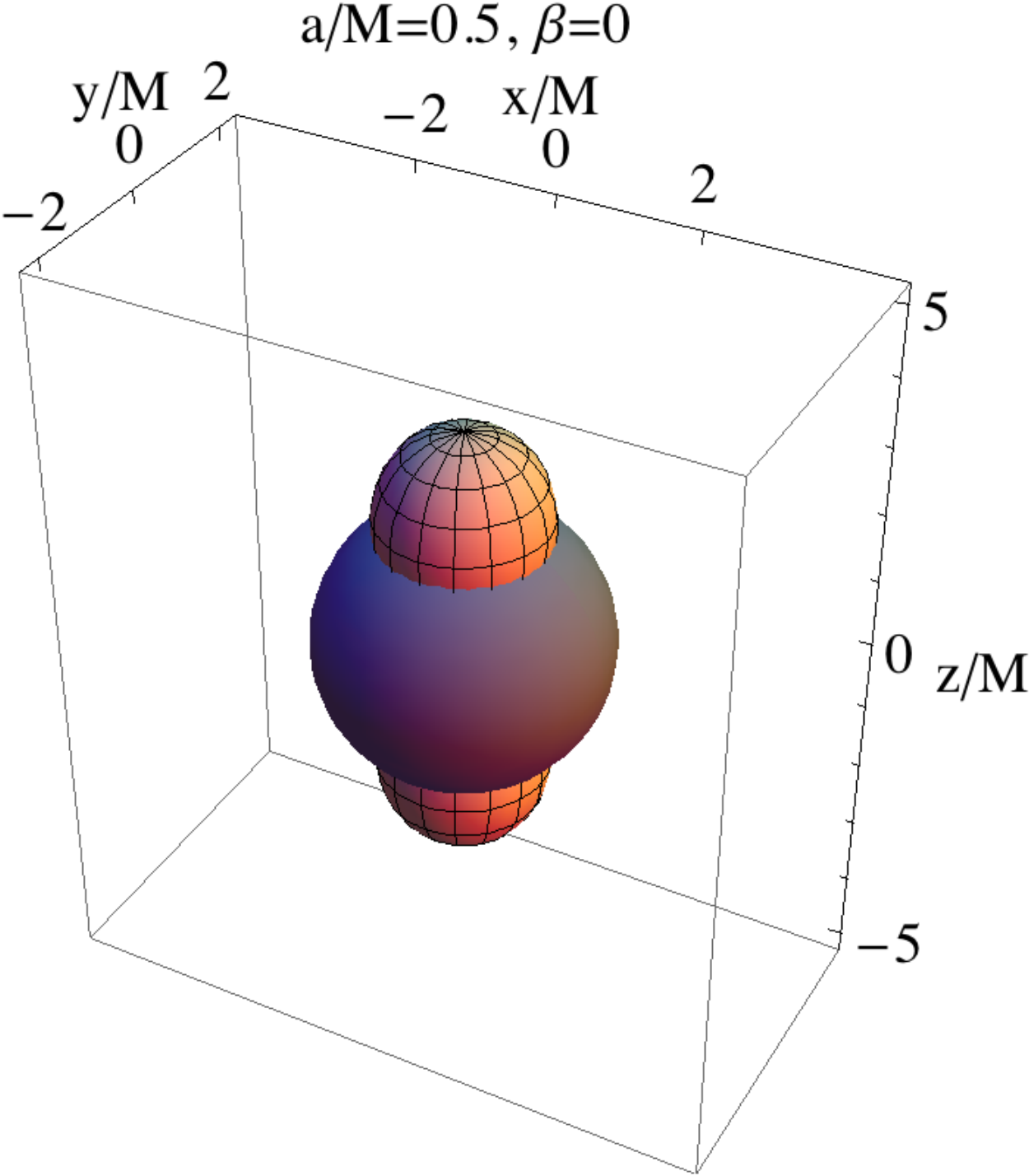}
\includegraphics[width=0.235\textwidth]{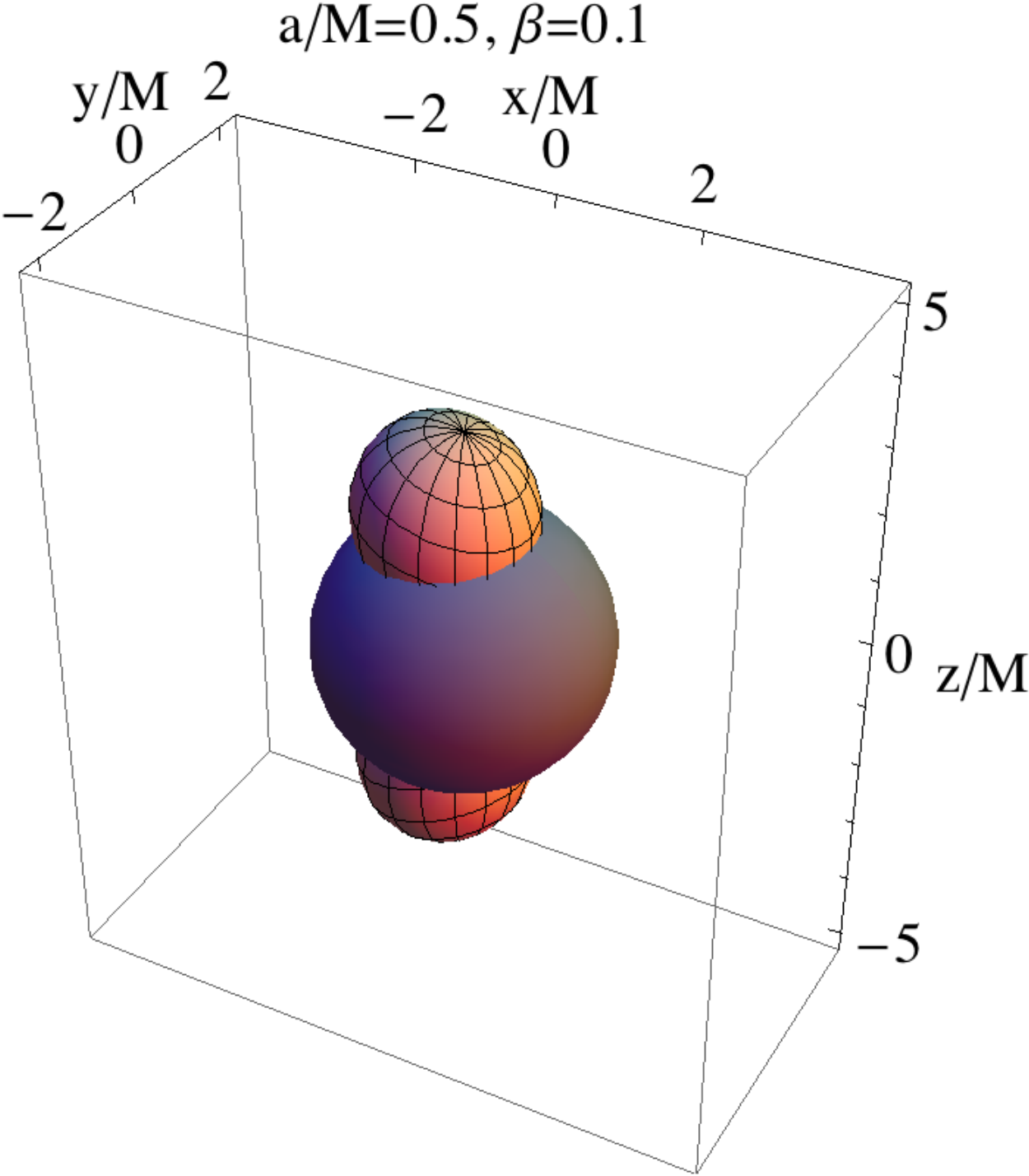}
\includegraphics[width=0.235\textwidth]{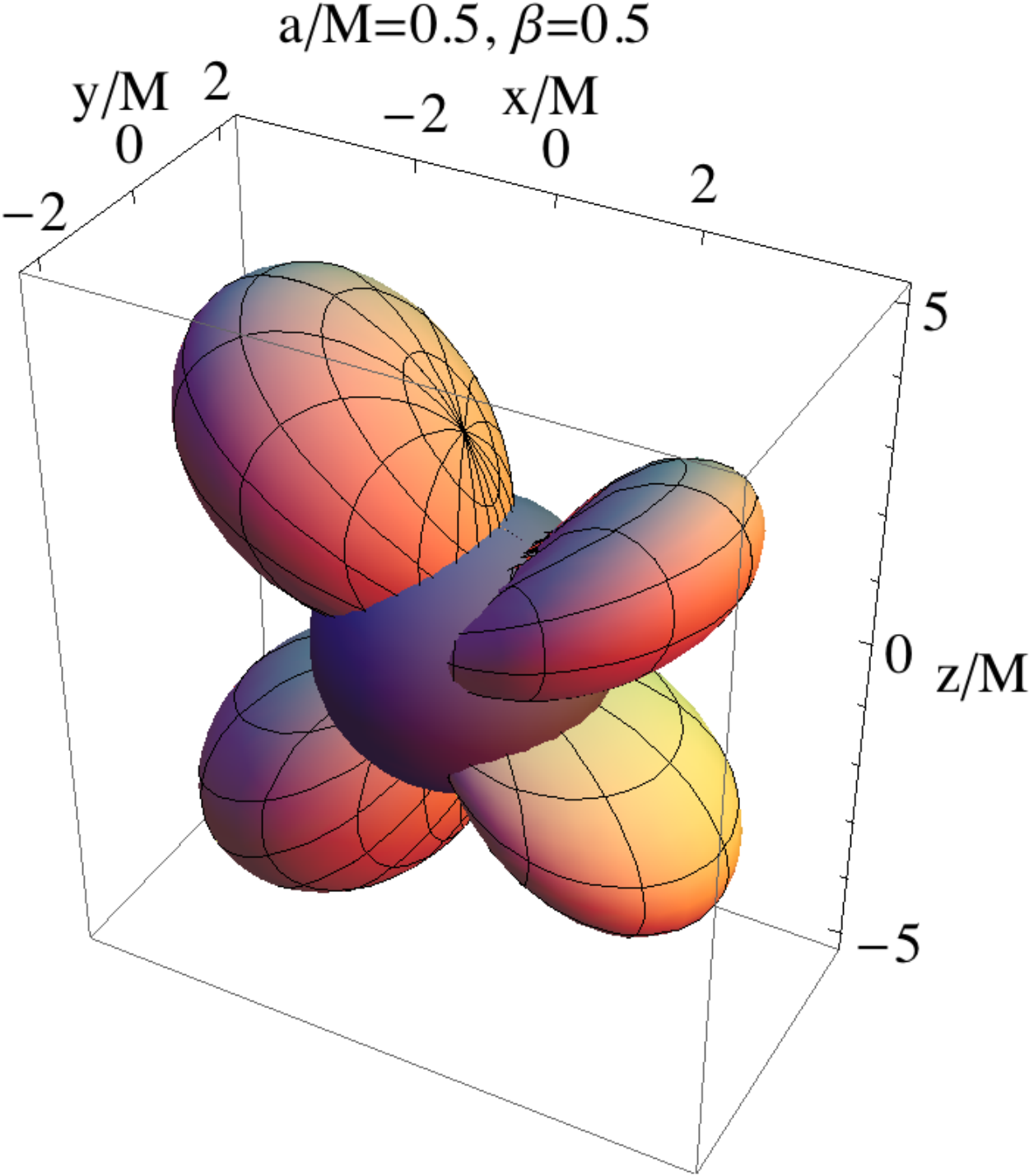}
\includegraphics[width=0.235\textwidth]{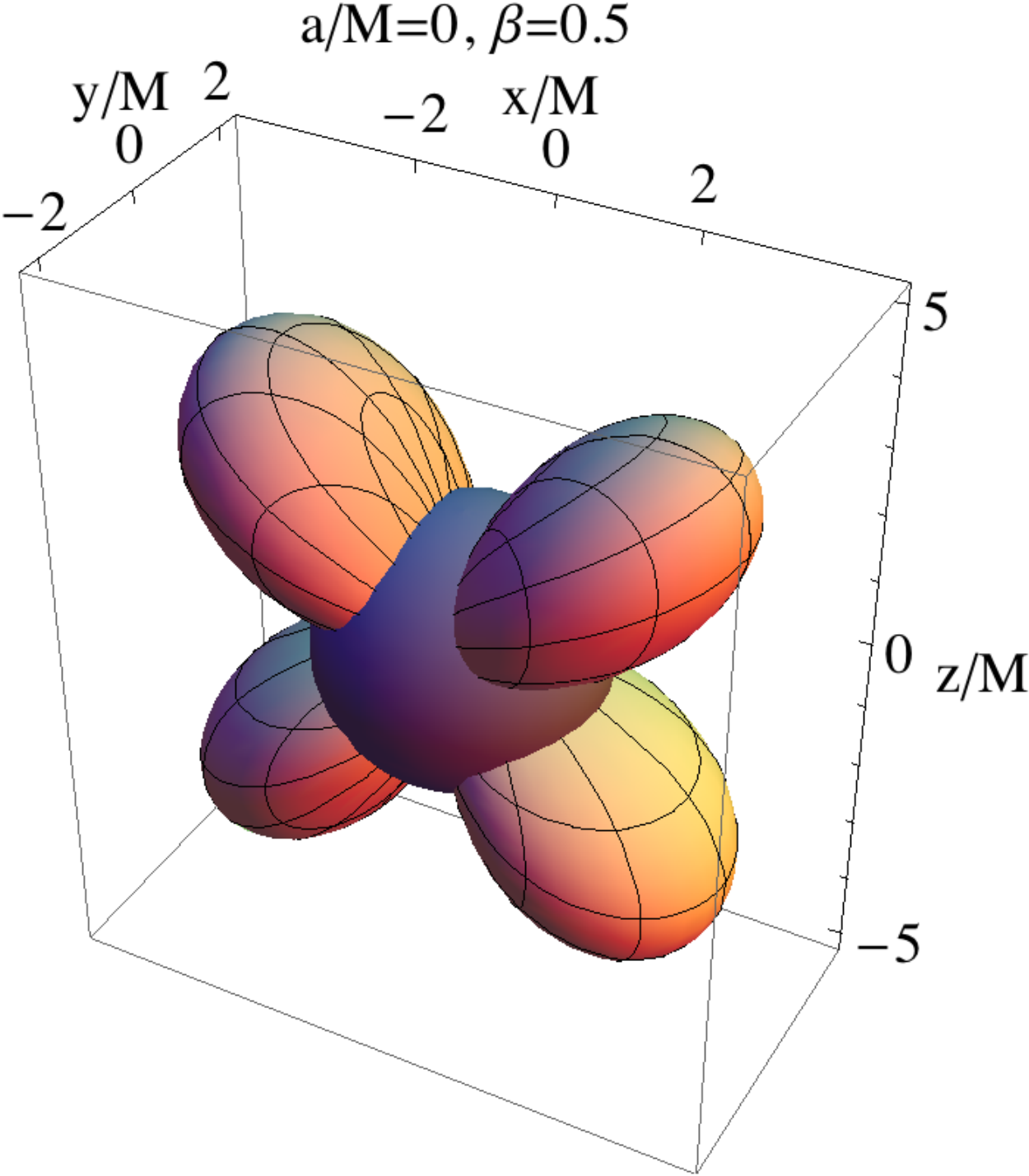}
\end{center}
\caption{Representative surfaces of the relativistic invariant
  $\boldsymbol{\vec{E}} \cdot \boldsymbol{\vec{B}}$. More specifically,
  the different panels show $\boldsymbol{\vec{E}} \cdot
  \boldsymbol{\vec{B}} = 0.1$ for different values of the dimensionless
  spin parameter $a/M$ and of the black hole dimensioneless speed
  $\beta$, as reported on the top of each panel. Shown as gray-shaded
  area is instead the black-hole event horizon.}
\label{Fig2}
\end{figure*}

\begin{figure*}
\begin{center}
\includegraphics[width=0.235\textwidth]{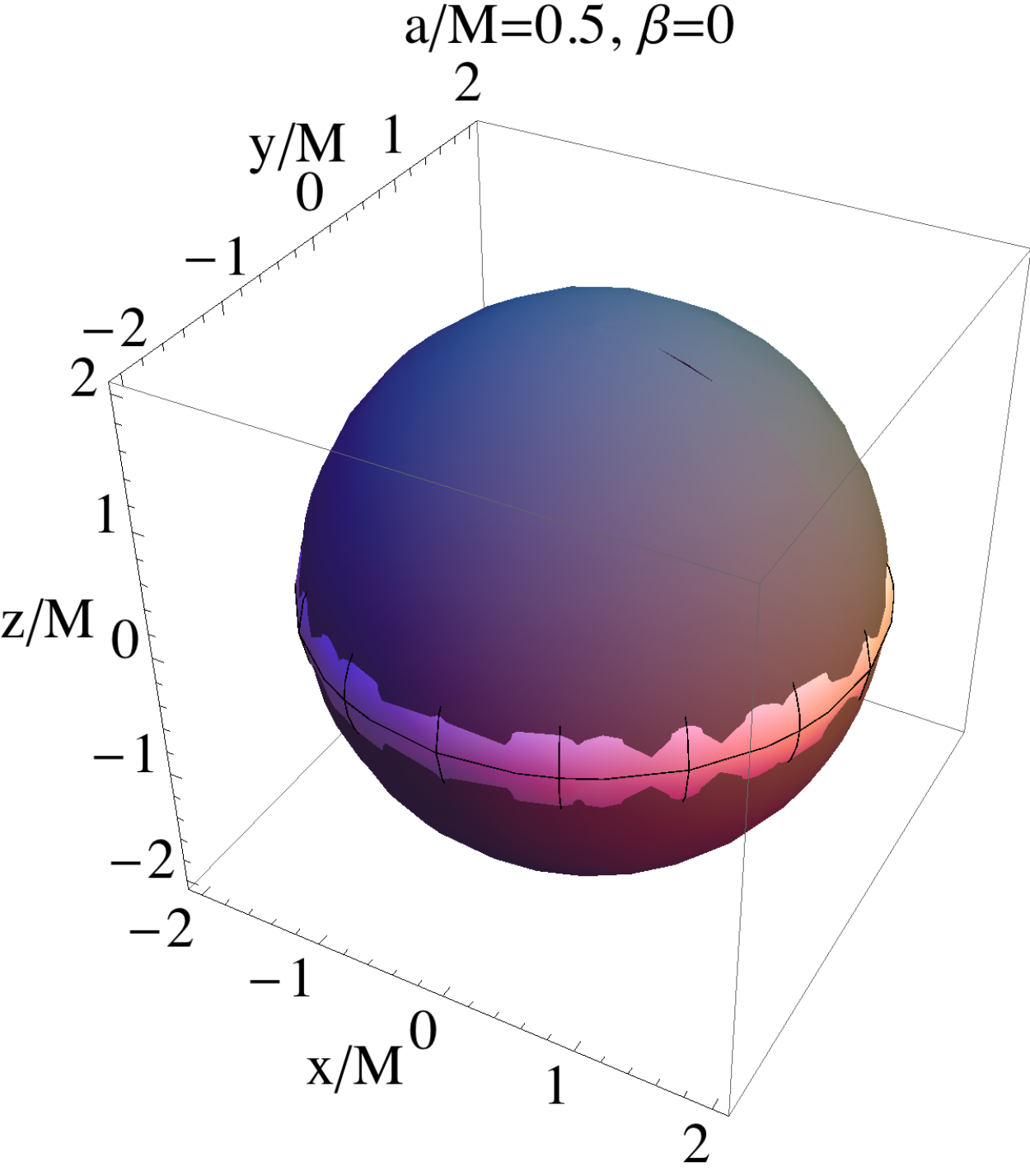}
\includegraphics[width=0.235\textwidth]{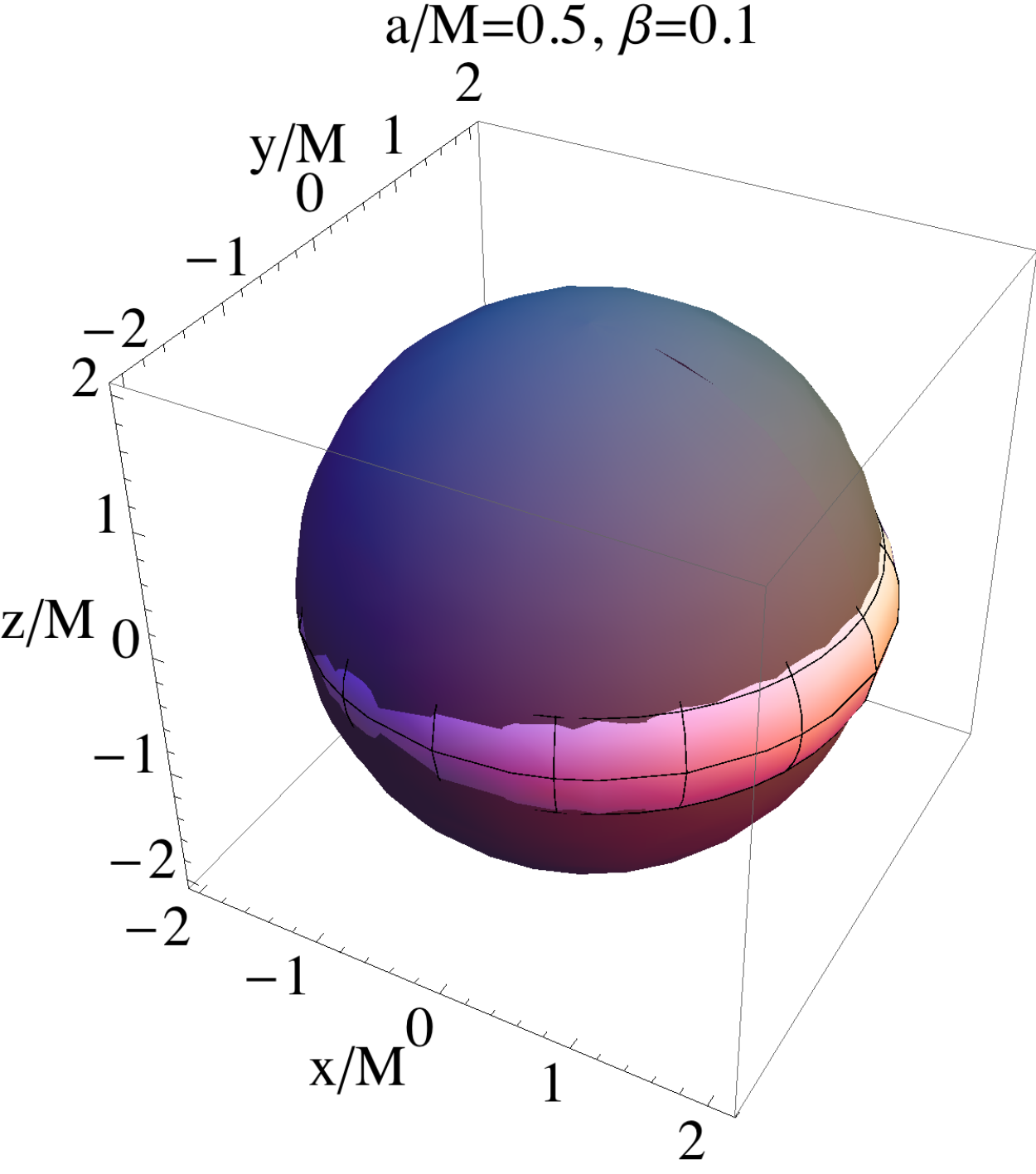}
\includegraphics[width=0.235\textwidth]{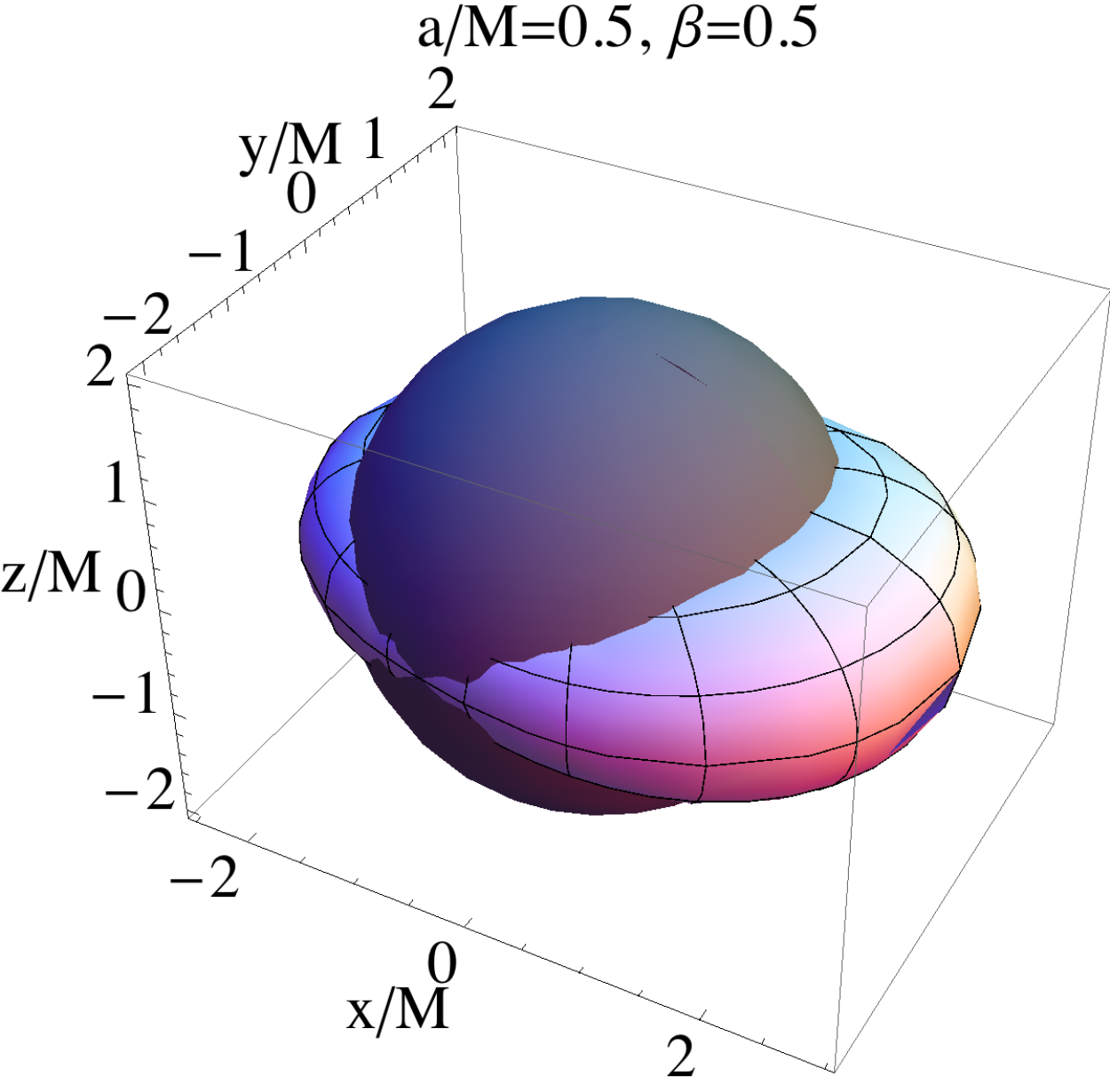}
\includegraphics[width=0.235\textwidth]{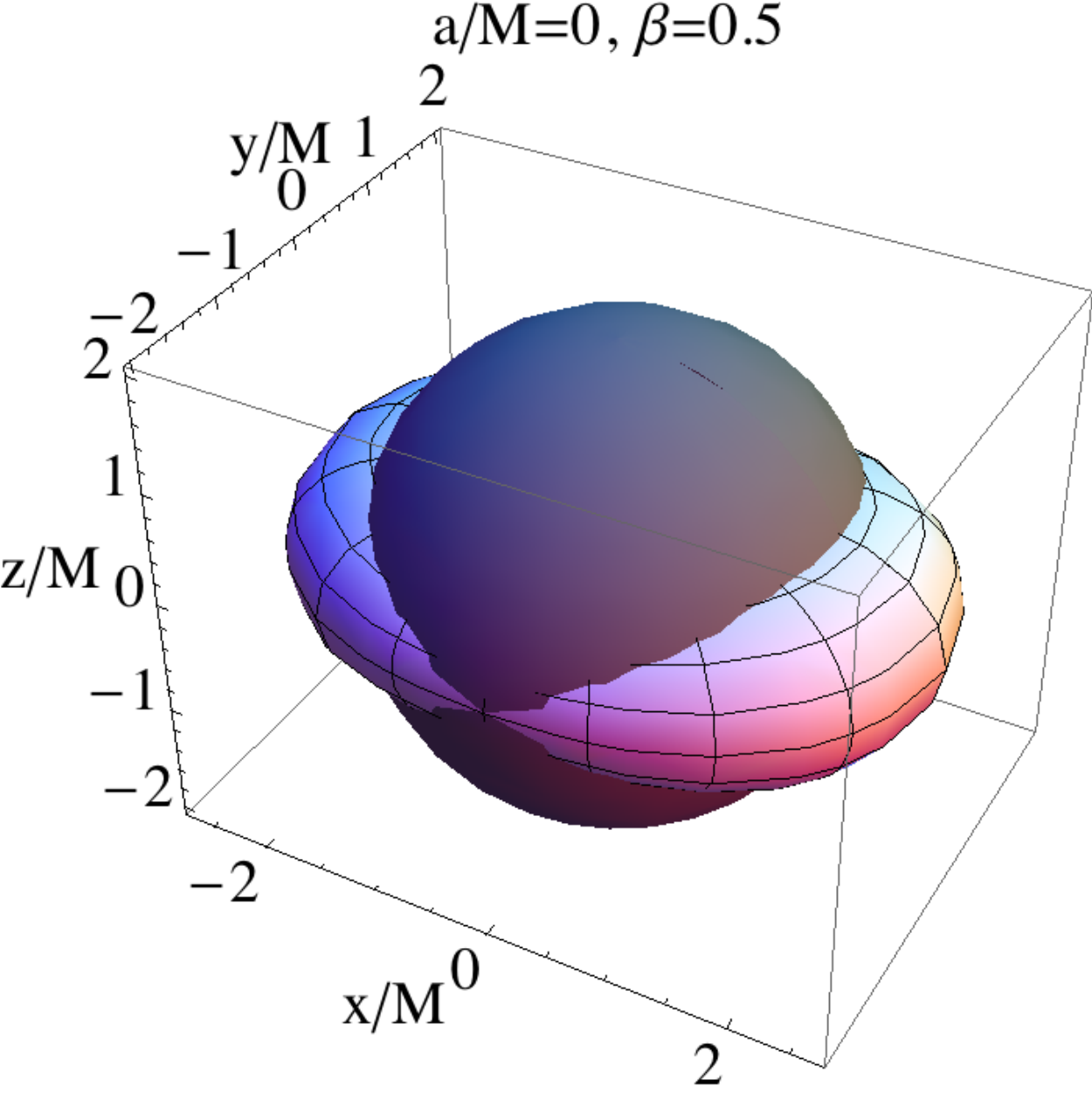}
\end{center}
\caption{The same as in Fig.~\ref{Fig2} but for surfaces at which
  $|\boldsymbol{\vec{E}}| = |\boldsymbol{\vec{B}}|$.} 
\label{Fig3}
\end{figure*}

Using the expressions~\eqref{eq:vecpot_boostedkerr} for the vector
potential, we next derive the expressions for the magnetic and electric
field as measured by the ZAMO observers for a moving Kerr black hole at
first order in the spin. The starting point is the calculation of the
nonzero components of the Faraday tensor, which a bit of algebra reveal
to be
\begin{eqnarray}
F_{rt}&=&A_{t,r}=\frac{aB_0M}{r^2}(\sin^2\theta-2)+E_0\sin\theta\cos\phi\,,
\\
F_{\theta t}&=&A_{t,\theta}=-\frac{2MaB_0}{r}\sin\theta\cos\theta
+E_0r\tilde{N}^{2}\cos\theta\cos\phi\,, 
\nonumber \\ \\
F_{\phi t}&=&A_{t,\phi}-A_{\phi,\,t}=-N^2E_0r\sin\theta\sin\phi\,,
\\
F_{\theta r}&=&A_{r,\theta}=-\tilde{N}^{2}aB_0\sin\theta\cos\theta
-2ME_0\cos\theta\cos\phi\,,
\nonumber \\ \\
F_{r \phi}&=&A_{\phi,r}=rB_0\sin^2\theta-2ME_0\sin\theta\sin\phi \,,
\nonumber \\ \\
F_{\theta \phi}&=&A_{\phi,\theta}=B_0\sin\theta\cos\theta r^2\,,
\end{eqnarray}
where we have retained terms $\mathcal{O}(a)$ and $\mathcal{O}(\beta)$,
neglecting terms $\mathcal{O}(a^2), \mathcal{O}(\beta^2)$ and
$\mathcal{O}(a \beta)$.

The physical components of the electric and magnetic fields are again
obtained after projecting on the ZAMOs tetrads and are given by
\begin{eqnarray}
\label{BKerr} 
&&B_{\hat{r}}=-B_0\cos\theta \,, \\ 
&&B_{\hat{\theta}}=\frac{1}{\tilde{N}}
\left(B_0\sin\theta-\frac{2ME_0}{r}\sin\phi\right)\,, \\
&&B_{\hat{\phi}}=-\frac{2E_0M}{r\tilde{N}}\cos\theta\cos\phi\
,
\end{eqnarray}
for the magnetic field (we recall that $E_0 \equiv \beta B_0$) and by
\begin{eqnarray}
\label{EKerr}
&&E_{\hat{r}}=E_0\sin\theta\cos\phi+\frac{aMB_0}{r^2}(3\sin^2\theta-2)\,, \\ &&
E_{\hat{\theta}}=\frac{1}{\tilde{N}}E_0\cos\theta\cos\phi\,, \quad \\
&&E_{\hat{\phi}}=\frac{1}{\tilde{N}}
\left(-E_0\sin\phi+\frac{2M}{r}B_0\sin\theta\right)\,, 
\end{eqnarray}
for the electric field. Similarly, the expressions for the electromagnetic field
invariants $E^2-B^2$ and $\boldsymbol{\vec{E}} \cdot
\boldsymbol{\vec{B}}$ take the form
\begin{widetext}
\begin{align}
\label{inv2}
\boldsymbol{\vec{E}}\cdot\boldsymbol{\vec{B}}=
\frac{2M}{r}B_0\cos\theta\left(E_0\sin\theta\cos\phi
-\frac{a}{4r}B_0(1+3\cos2\theta)\right)\,, 
\end{align}
\begin{align}
\label{inv1} B^2-E^2 =  B_0^2\Bigg\{ & (1-\beta^2)
+\frac{2M}{r}\left[\beta^2(\cos^2\theta\cos^2\phi+\sin^2\phi)-\sin^2\theta\right]+\frac{a\beta
  M}{2r^2} \left(3\sin3\theta-\sin\theta\right)\cos\phi \nonumber \\ &
+\frac{a^2}{4r^2}
\left[2(\cos2\theta-1)+\frac{M}{r}(\cos4\theta-1) + 
\frac{M^2}{2r^2}(1-28\cos2\theta-5\cos4\theta)\right]\Bigg\}\ ,
\end{align}
\end{widetext}
while the expressions for the parallel electric field
$\boldsymbol{\vec{E}}_{\parallel}$ and for the charge density
$\rho_{\rm ind}$ can be calculated to be respectively
\begin{widetext}
\begin{eqnarray}
\label{eq:Epar}
&&E_{\parallel}=\frac{\tilde{N}M}{r\sqrt{1+{2M}/{r}\cos^2\theta}}
\Bigg[2E_0\sin\theta\cos\theta\cos\phi
-\frac{a}{2r}B_0\cos\theta(1+3\cos2\theta)\Bigg]\,,
\end{eqnarray}
and
\begin{eqnarray}
\label{eq:rho_lin} 
\rho_{\rm ind}=\frac{B_0M}{32\pi r^3\tilde{N}
\left(1+ {2M \cos^2\theta }/{r}\right)^2} && \Bigg\{8\beta
r\cos\phi\sin\theta\Bigg[1+\frac{7M^2}{r^2}+\frac{6M}{r}
\nonumber
\\&& \hskip -3.0cm
+\left(\frac{2M}{r}+1\right)\left(\frac{4M}{r}+3\right)\cos2\theta
+\frac{M^2}{r^2}\cos4\theta\Bigg]
-a\Bigg[5+\frac{24M}{r}+\frac{26M^2}{r^2}
 \nonumber
\\ +\left(\frac{5M}{r}+3\right)\left(\frac{11M}{r}+4\right)\cos2\theta
&& +\left(\frac{38M^2}{r^2}+\frac{48M}{r}+15\right)\cos4\theta
+\frac{3M}{r}\left(\frac{3M}{r}+1\right)\cos6\theta\Bigg]\Bigg\}\,.
\end{eqnarray}
\end{widetext}
Note that in addition to the lowest-order terms, the invariant
$B^2-E^2$ also contains terms $\mathcal{O}(a^2)$, and
$\mathcal{O}(\beta^2)$, since no terms $\mathcal{O}(a)$, or
$\mathcal{O}(\beta)$ are present.

\begin{figure*}
\includegraphics[width=0.235\textwidth]{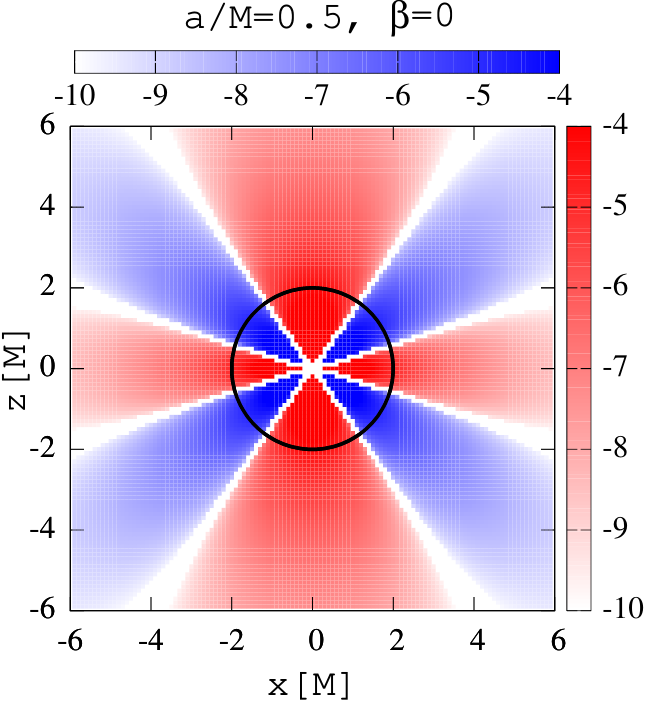}
\includegraphics[width=0.235\textwidth]{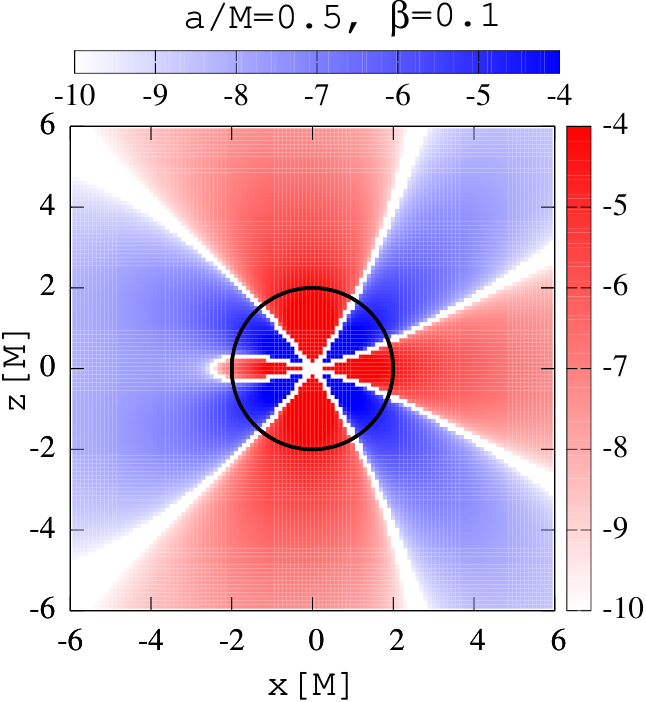}
\includegraphics[width=0.235\textwidth]{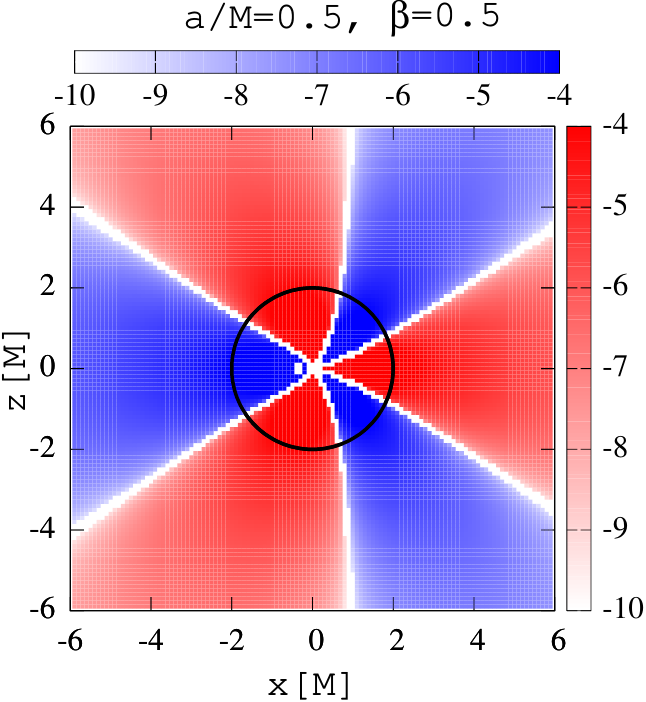}
\includegraphics[width=0.235\textwidth]{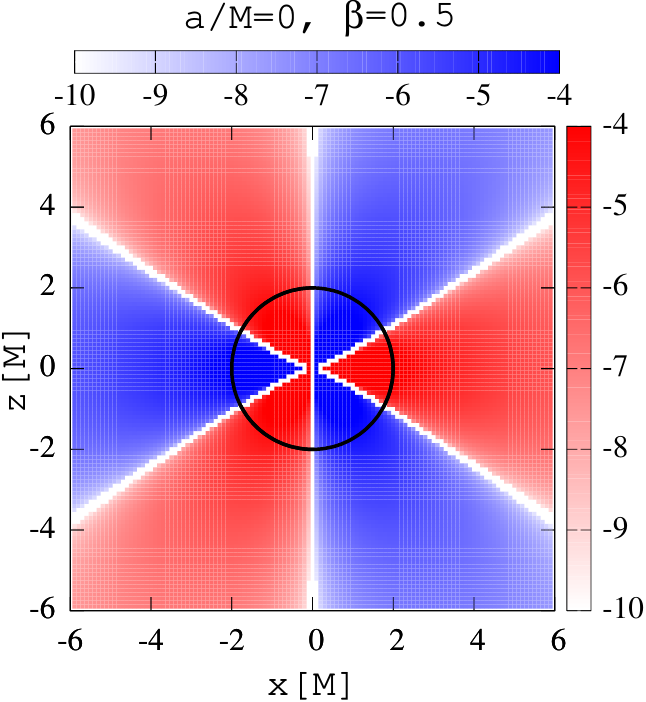}
\vskip 0.5cm
\includegraphics[width=0.235\textwidth]{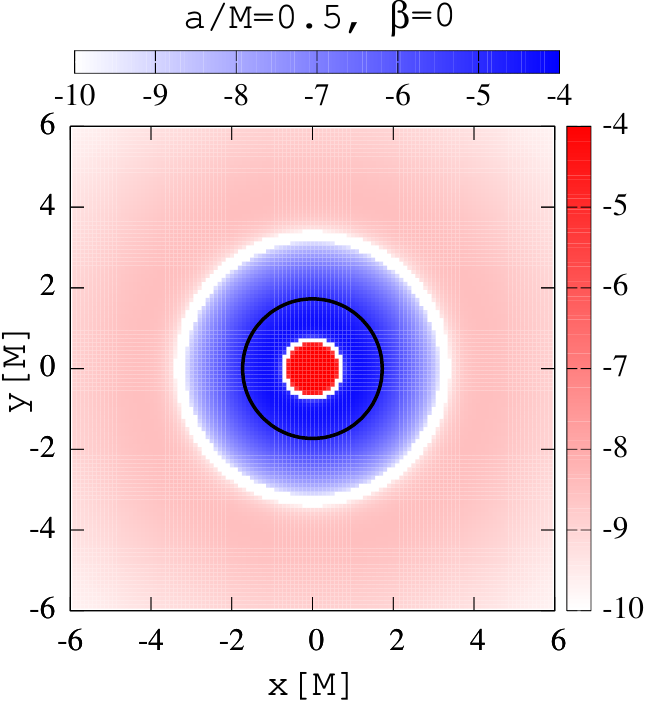}
\includegraphics[width=0.235\textwidth]{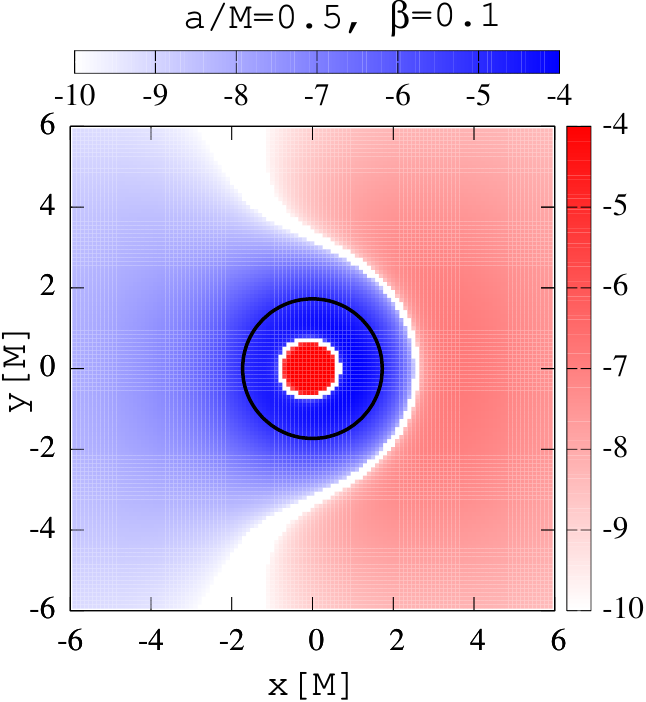}
\includegraphics[width=0.235\textwidth]{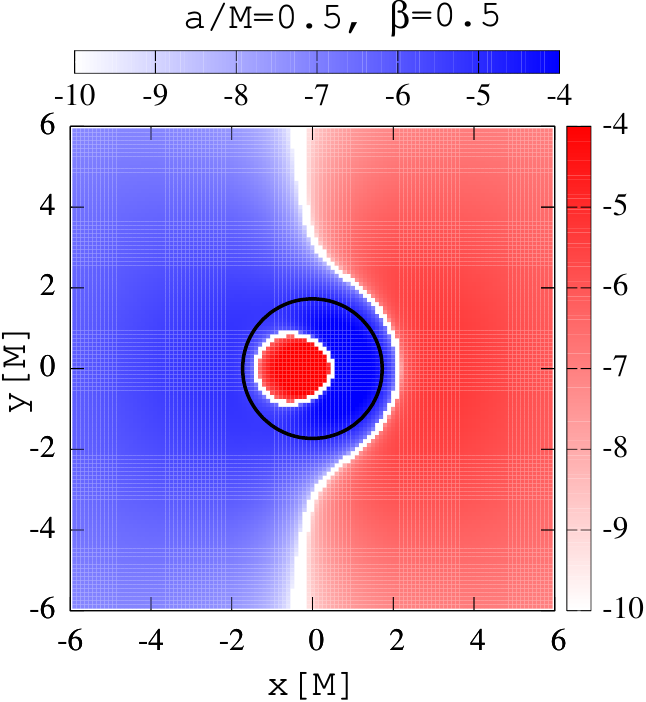}
\includegraphics[width=0.235\textwidth]{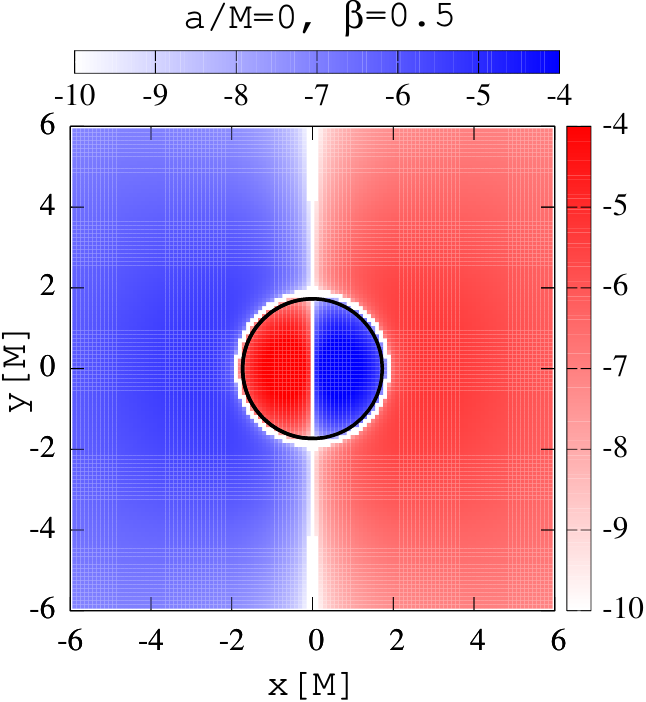}
\caption{Induced charge-density distribution $\rho_{\rm ind}$ for $M=1$,
  $B_0=1$, and different values of $a/M$ and $\beta$. The charge
  distribution is indicated with a colorscale in which blue refers to
  positive charges and red to negative charges, respectively. The upper
  row refers to a projection in the $(x,z)$ plane with a logarithmic
  colorscale, while the bottom one to a projection in $(x,y)$ plane at
  $z/M=1$. Indicated with a solid black circles are the black hole
  horizons. Regions in white correspond to zero induced charges and thus
  separate two domains with opposite charges.}
\label{Fig3b}
\end{figure*}
\begin{figure}
\includegraphics[width=0.235\textwidth]{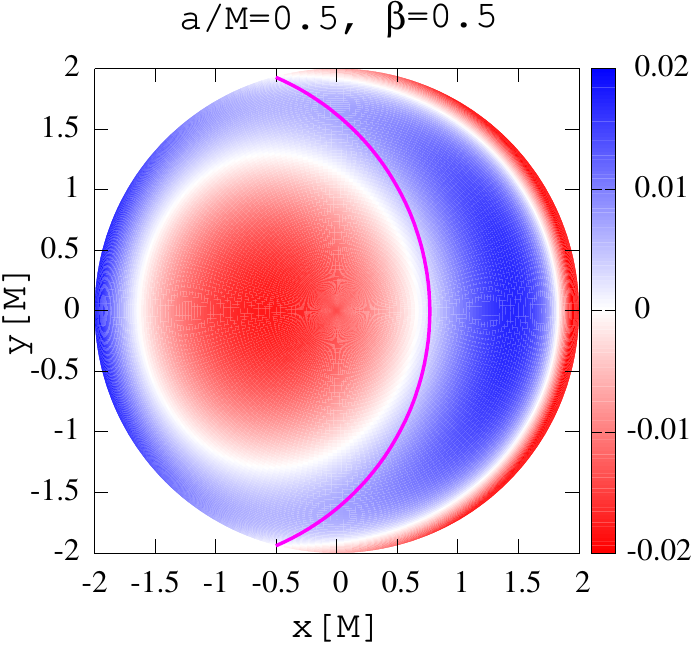}
\includegraphics[width=0.235\textwidth]{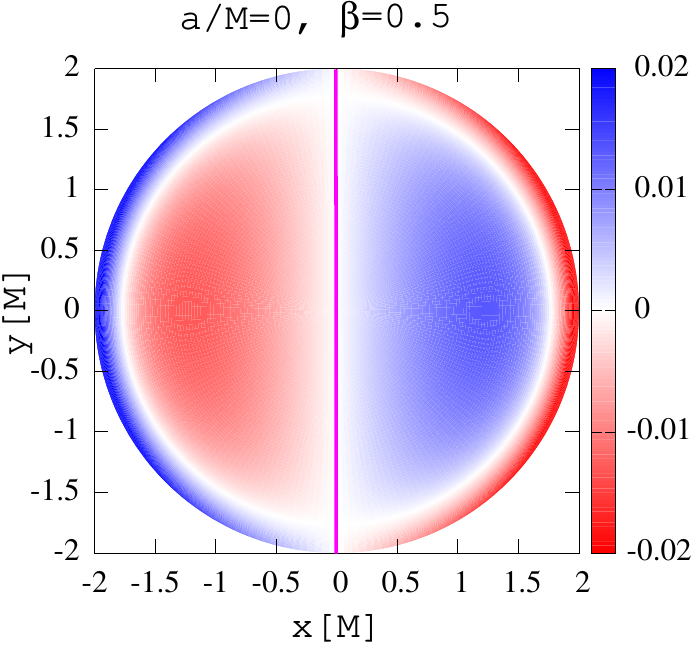}
\caption{Projection on the $(x,y)$ plane of the induced charge-density
  distribution on the event horizon. The colormaps are the same as in
  Fig.~\ref{Fig3b} but shown with a linear colorscale to highlight the
  small-scale variations. The two panels refer to the last two columns in
  Fig.~\ref{Fig3b} and we have indicated with magenta solid line
  separatrixes $E_{\parallel}=0$.}
\label{Fig3c}
\end{figure}

Figure~\ref{Fig2} shows representative surfaces of the relativistic
invariant $\boldsymbol{\vec{E}} \cdot \boldsymbol{\vec{B}}$ and, more
specifically, the surface $\boldsymbol{\vec{E}} \cdot
\boldsymbol{\vec{B}} = 0.1$. The different panels refer to different
values of the spin parameter $a/M$ and of the black hole velocity
$\beta$, as reported on the top of each panel. Shown instead as
gray-shaded area is the black-hole event horizon, which is located at
$2M$ [the true position of the horizon is of course $r=M+\sqrt{M^2-a^2}
  \simeq 2M + \mathcal{O}(a^2)$]. Note that although the axial symmetry
is broken by the presence of a velocity, reflection symmetries are still
present with respect to the $(x,z)$ plane and to the $(x,y)$
plane. Similarly, Fig.~\ref{Fig3} shows the position of the surfaces at
which $|\boldsymbol{\vec{E}}| = |\boldsymbol{\vec{B}}|$ for the same
values of the parameters $a/M$ and $\beta$ selected in
Fig.~\ref{Fig2}. Note that the reflection symmetry across the $(y,z)$
plane is lost as soon as the black hole is rotating (\ie for $a \neq 0$),
but that the reflection symmetries with respect to the $(x,z)$ and
$(x,y)$ planes remain.

Finally, in Fig.~\ref{Fig3b} we show the induced charge density
$\rho_{\rm ind}$ for the same set of parameters $a/M$ and $\beta$ used in
the previous two figures. More specifically, the upper row refers to a
projection in the $(x,z)$ plane and where we have indicated with a solid
black line the black-hole horizon. The bottom row is the same as the top
one but with a projection in $(x,y)$ plane and we have indicated with a
magenta solid line the separatrix $E_{\parallel}=0$. Note that in all
panels the areas in white correspond to regions with zero induced charges
and thus separate two domains with opposite charges (we have indicated
with blue regions of positive charges and with red regions of negative
charges). Note also that the velocity-induced corrections are generically
the most important ones, with the charge density assuming its maximum
values in the equatorial plane and that the spin-induced corrections are
evident only for comparatively small values of the velocity (cf. the
first two panels of Fig.~\ref{Fig3b}. Finally, we note that the use of
the Kerr-Schild coordinates allows us to have a complete and regular
description of the charge distribution even inside the event horizon, in
contrast with the results of~\cite{Lyutikov:2011}, where the distribution
was singular at the horizon. This is highlighted in Fig.~\ref{Fig3c},
which shows a magnification of the projections on the $(x,y)$ plane of
the induced charge-density distribution on the event horizon. The
colormaps are the same as in Fig.~\ref{Fig3b} but shown with a linear
colorscale to highlight the small-scale variations. The two panels refer
to the last two columns in Fig.~\ref{Fig3b} and we have indicated with
magenta solid line separatrixes $E_{\parallel}=0$. An additional
discussion of this figure will be made also in the following Section.

\section{electromagnetic energy losses}
\label{LOSS}

In the scenario considered here, all of the magnetic field lines
threading the horizon of the moving and rotating black hole are connected
to infinity. Under these conditions, winds of charged particles can be
produced and flow along the magnetic field lines, being accelerated by
the electric-field components parallel to the magnetic-field lines. The
origin of the particles in the wind is still a matter of debate, but as
discussed in Ref.~\cite{Lyutikov:2011}, it is possible that the potential
gap in the vicinity of the black hole is large enough for the creation of
particles due to electron-positron cascades. Alternatively, any stray
charged particle orbiting near the black hole could be accelerated to
high Lorentz factors by the large electric fields and thus be able to
produce pairs.

If the acceleration of charged particles and a cascade via pair
generation takes place in the vicinity of the black hole, it may build up
a plasma magnetosphere. Obviously, the charge density and the velocity
distribution of the magnetospheric plasma cannot be described accurately
within our approach, but we can exploit the analogies with the physics of
pulsar magnetosphere. We can therefore estimate $\rho_{\rm ind}$ as the
induced charge density needed in order to screen the parallel component
of electric field\footnote{Of course, the charge density required
    to screen the parallel component of the electric field is $-\rho_{\rm
      ind}$, but hereafter we will not pay attention to the sign as we
    are interested in the absolute values of the currents flowing in the
    magnetosphere}.  Furthermore, assuming that the charged particles
are rapidly accelerated up to relativistic velocities, we can take as
modulus of the current density in the black hole magnetosphere the upper
limit $c\rho_{\rm ind}$. These estimates are clearly very crude and
analogous to those made in Ref.~\cite{Lyutikov:2011}, but are useful as
they provide us with a bulk reference current through which we can
estimate the electromagnetic energy losses within the membrane
paradigm~\cite{Price86a}. As we will discuss later on, a comparison with
numerical-relativity calculations shows that the error associated to this
approximation is only of a factor of a few.

We recall that the membrane paradigm of black-holes electrodynamics
represents an elegant and very useful approach in which relativistic
black-hole electrodynamics resembles classical electrodynamics
\cite{Price86a}. In particular, the membrane paradigm considers the
extended horizon of the black hole as a conducting sphere which is
endowed with physical properties, such as a surface resistivity, a
surface charge and a current density. Although this is only an effective
construction, the membrane paradigm serves as a useful physical
guideline, providing the interpretation of the complex phenomena taking
place in the vicinity of black holes (see~\cite{Penna2013} for a recent
comparison between the membrane-paradigm description of the
Blandford-Znajek mechanism and the results of general-relativistic
magnetohydrodynamical simulations).

In order to estimate the electromagnetic energy losses it is important to
know the distribution of the charges and currents in and out of the
black-hole magnetosphere and, in particular, the location of the surfaces
where $E_{\parallel}=0$ or $\rho_{\rm ind}=0$. These surfaces, in fact,
divide the space in domains where electric currents of different sign are
nonzero and of variable intensity. As a result, the two-dimensional
surfaces with $E_{\parallel}=0$ and $\rho_{\rm ind}=0$ mark the
separation between regions with currents in different directions and
charges of opposite signs, respectively. Such surfaces are simply
obtained from Eqs.~\eqref{eq:Epar} and \eqref{eq:rho_lin} and thus
described by the following equations
\begin{equation}
\label{Epareq0} 4\beta r\sin\theta\cos\phi-{a}(1+3\cos2\theta)=0\,, 
\qquad \cos\theta=0\,,
\end{equation}
and
\begin{widetext}
\begin{eqnarray}
\label{rhoindeq0} && 8\beta
r\cos\phi\sin\theta\Bigg[1+\frac{7M^2}{r^2}+\frac{6M}{r}
+\left(\frac{2M}{r}+1\right)\left(\frac{4M}{r}+3\right)\cos2\theta
+\frac{M^2}{r^2}\cos4\theta\Bigg]
-a\Bigg[5+\frac{24M}{r}+\frac{26M^2}{r^2} \nonumber \\ &&
\quad+\left(\frac{5M}{r}+3\right)\left(\frac{11M}{r}+4\right)\cos2\theta
+\left(\frac{38M^2}{r^2}+\frac{48M}{r}+15\right)\cos4\theta
+\frac{3M}{r}\left(\frac{3M}{r}+1\right)\cos6\theta\Bigg]=0\,.
\end{eqnarray}
\end{widetext}

In the limiting case when $a=0, \ \beta \neq 0$, the first of
Eqs. (\ref{Epareq0}) divides the space into four quadrants, while the
second one simply marks the equatorial plane (\ie $\cos\theta=0$);
similarly, when $a \neq 0, \ \beta = 0$, the first of
Eqs. (\ref{Epareq0}) marks conical surfaces in the upper and lower
hemispheres (\ie $1+3\cos2\theta=0$). In the more general case of nonzero
values for $a$ and $\beta$, the surface with $E_{\parallel}=0$ is rather
complex, but it represents, overall, the transition between these two
limiting cases. It should be noted that the different scaling with radius
of the spin-induced contributions (that decrease like $1/r^3$) and of the
velocity-induced contributions (that decrease like $1/r^2$) is such that
the former are important only in the vicinity of the black hole. At large
distances, therefore, the shape of the separating surfaces will be quite
similar to the case of nonrotating black holes discussed in
Ref.~\cite{Lyutikov:2011}.

\begin{figure*}
\begin{center}
\includegraphics[width=0.235\textwidth]{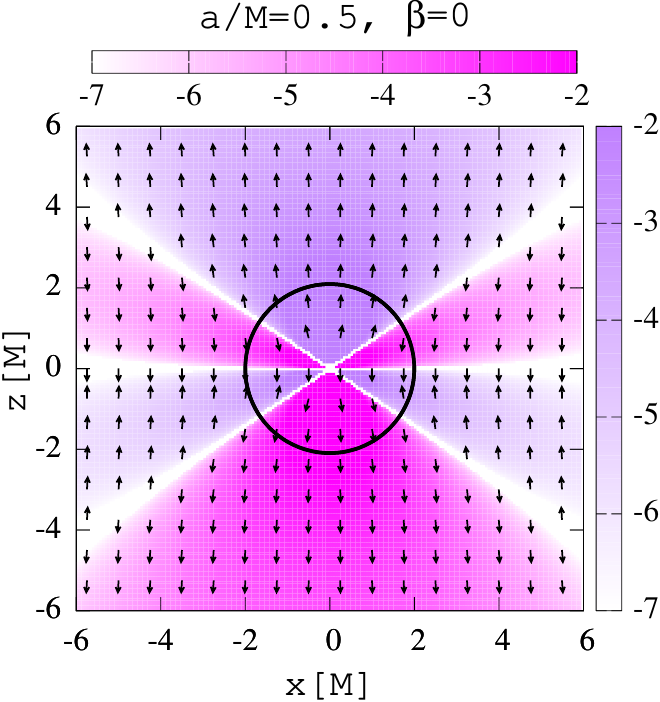}
\includegraphics[width=0.235\textwidth]{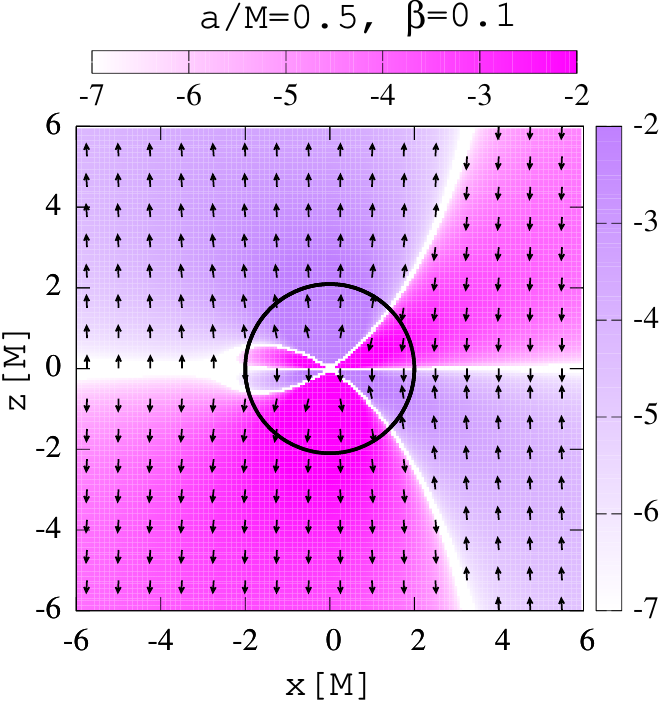}
\includegraphics[width=0.235\textwidth]{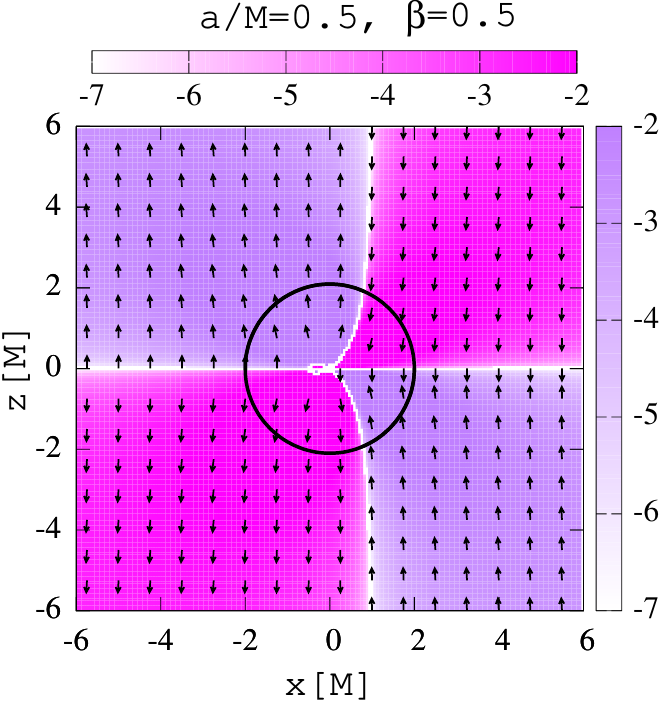}
\includegraphics[width=0.235\textwidth]{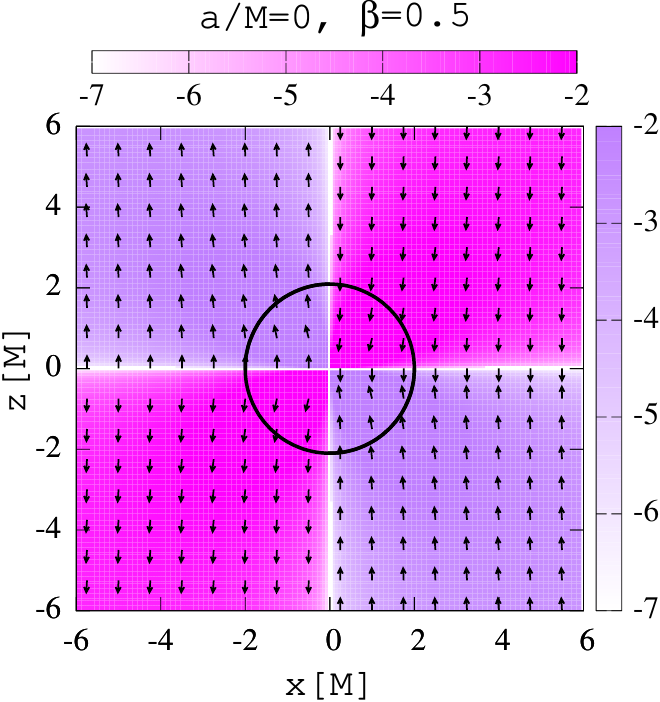}
\end{center}
\caption{Electric-field component parallel to the magnetic field,
  $E_{\parallel}$ in the $(x,z)$ plane. The arrows show the direction of
  the field lines, while the solid black circle marks the position of the
  horizon. Violet colors show the regions where the considered component
  of the electric field is directed along the magnetic field lines, while
  the magenta colors show the opposite.}
\label{Fig4}
\end{figure*}

\begin{figure}
\begin{center}
  \includegraphics[width=0.23\textwidth]{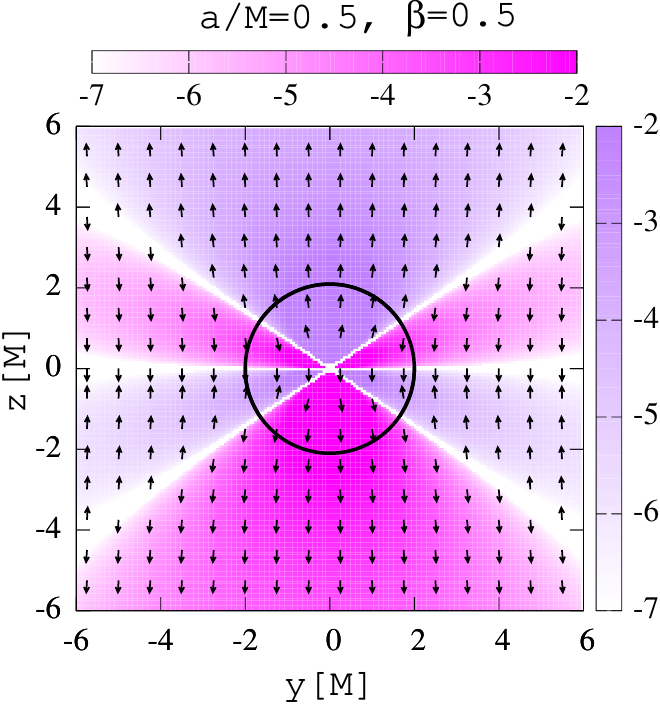}
  \includegraphics[width=0.23\textwidth]{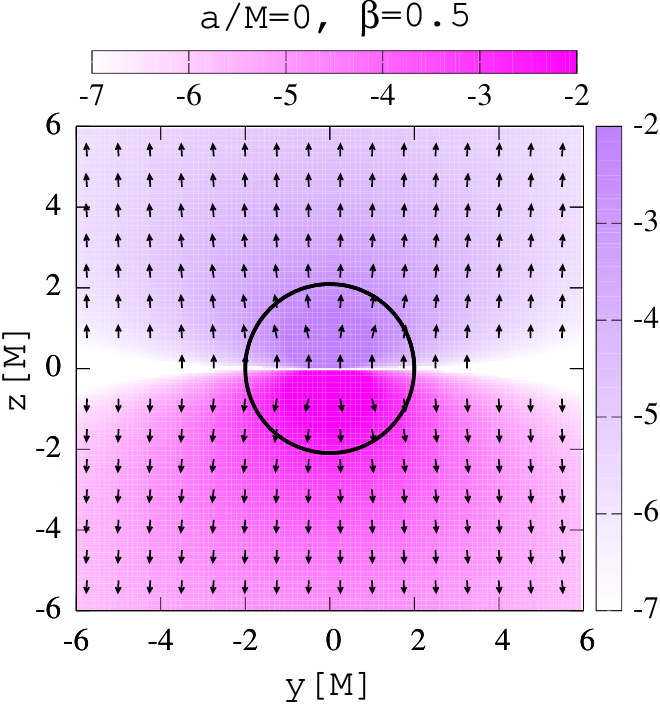}
\end{center}
\caption{The same as in Fig.~\ref{Fig4}, but for the electric field in
  the $(y,z)$ plane. Note that we report only the solutions corresponding
  to the last two panels of Fig.~\ref{Fig4}, as these are the most
  representative ones, with the solution in this plane being essentially
  insensitive to changes in the velocity.} \label{Fig5}
\end{figure}

Figure~\ref{Fig4} reports the electric field $E_{\parallel}$ in the
$(x,z)$ plane and for the same values of the parameters $a$ and $\beta$
used in the previous figures. More specifically, the arrows show the
direction of the parallel electric field lines, while the black solid
line marks the position of the horizon. Violet colors refer to regions
where the electric field is in the same direction as the magnetic field
lines, while the magenta colors show the opposite; in either case, the
reported colormaps can be used to deduce the strength of the field. As
remarked for the induced charge density, also the induced parallel
electric field is most sensitive to the velocity of the black hole and
the qualitative features of the solution do not change appreciably for
$\beta \gtrsim 0.5$. This is more evident for the electric field
distribution in the $(y,z)$ plane, which is shown in Fig.~\ref{Fig5}, and
where we report only the solutions corresponding to the last two panels
of Fig.~\ref{Fig4}; these field structures are essentially insensitive to
further increases in the velocity.

We should note that when considering the $x=0$ plane, \ie for
$\phi=\pi/2$, the first of Eqs. (\ref{Epareq0}) loses its dependence on
$\beta$ and thus the position of the separating surfaces
$E_{\parallel}=0$ in this plane remains the same as in the left panel of
Fig.~\ref{Fig5} for any nonzero $a$ and any $\beta$. Furthermore, when
$a=0$, the $x=0$ plane contains the separatrix and thus $E_{\parallel}=0$
on this plane. In order to show a nonzero parallel field, the right panel
of Fig.~\ref{Fig5} refers to a $(y,z)$ plane at $x=0.3\,M$.

For arbitrary values of $a$ and $\beta$, the position of the separatrixes
will be such that each hemisphere will be divided on four domains with
different directions of the electric currents.  In order to estimate the
energy losses within a membrane-paradigm approach, it is necessary to
have the values of the currents close to the horizon. In contrast with
Ref.~\cite{Lyutikov:2011}, this is not problematic to do here, since our
use of Kerr-Schild coordinates implies that we can consider and integrate
necessary quantities directly at the horizon, as none of the equations
discussed in the previous sections is singular there. In particular, at
the horizon the induced charge density is equal to
\begin{widetext}
\begin{eqnarray}
\label{rhohor} &&\rho_{\rm ind}=\frac{B_0}{32\sqrt{2}\pi
M}\frac{1}{(1+\cos^2\theta)^2} \Bigg[4\beta\cos\phi\sin\theta
  \left(\cos^4\theta+9\cos^2\theta-2\right)- \frac{a}{M}
\left(15\cos^6\theta+26\cos^4\theta-27\cos^2\theta+2\right)\Bigg]\,,
\nonumber \\ 
\end{eqnarray}
\end{widetext}
and the separatrixes are given by
\begin{equation}
\label{sepEhor}
\sin\theta=
-\frac{{2\beta M}\cos\phi - 
\sqrt{{4\beta^2 M^2}\cos^2\phi+6 {a^2}}}{3{a}}\,,
\end{equation}
and
\begin{eqnarray}
\label{sixorder} 
15 a \sin^6\theta &+& 
4\beta M \cos\phi\sin^5\theta-71a\sin^4\theta \\ \nonumber  
&& \hskip -1.5cm
-44\beta M \cos\phi\sin^3\theta 
+ 70a\sin^2\theta \\ \nonumber  
&& \hskip -0.5cm +32\beta M\cos\phi\sin\theta-16a=0\,.
\end{eqnarray}
Some examples of the position of separatrixes on the black-hole horizon
as projected on the $(x,y)$ plane are presented in Fig.~\ref{Fig3c}. The
separatrixes for the induced charge density are indicated with white,
while magenta lines indicate the position of the separatrix
$E_{\parallel}=0$. Note that in the left panel of Fig.~\ref{Fig3c}, one
of the separatrixes for $\rho_{\rm ind}$ intersects the equatorial plane
at the values of $\phi \sim \pi/2$ and $\phi \sim 3\pi/2$. These points
correspond to the solution of Eq.~(\ref{sixorder}) when $\theta=\pi/2$
and thus satisfy the condition
\begin{equation}
\label{phi12} 
\cos\phi=-\frac{a}{4\beta M}\,.
\end{equation}
Equation \eqref{phi12} has two distinct solutions only for $\beta \geq
a/4 M$; for velocities that are instead $\beta < a/4 M$, a separatrix for
the charge density $\rho_{\rm ind}$ still exists but is not contained in
the equatorial plane. As can be seen from Eq. (\ref{sepEhor}), and in the
left panel of Fig.~\ref{Fig3c}, the separatrix $E_{\parallel}=0$
intersects the equatorial plane in the same points.

\begin{figure}
\begin{center}
  \includegraphics[width=0.4\textwidth]{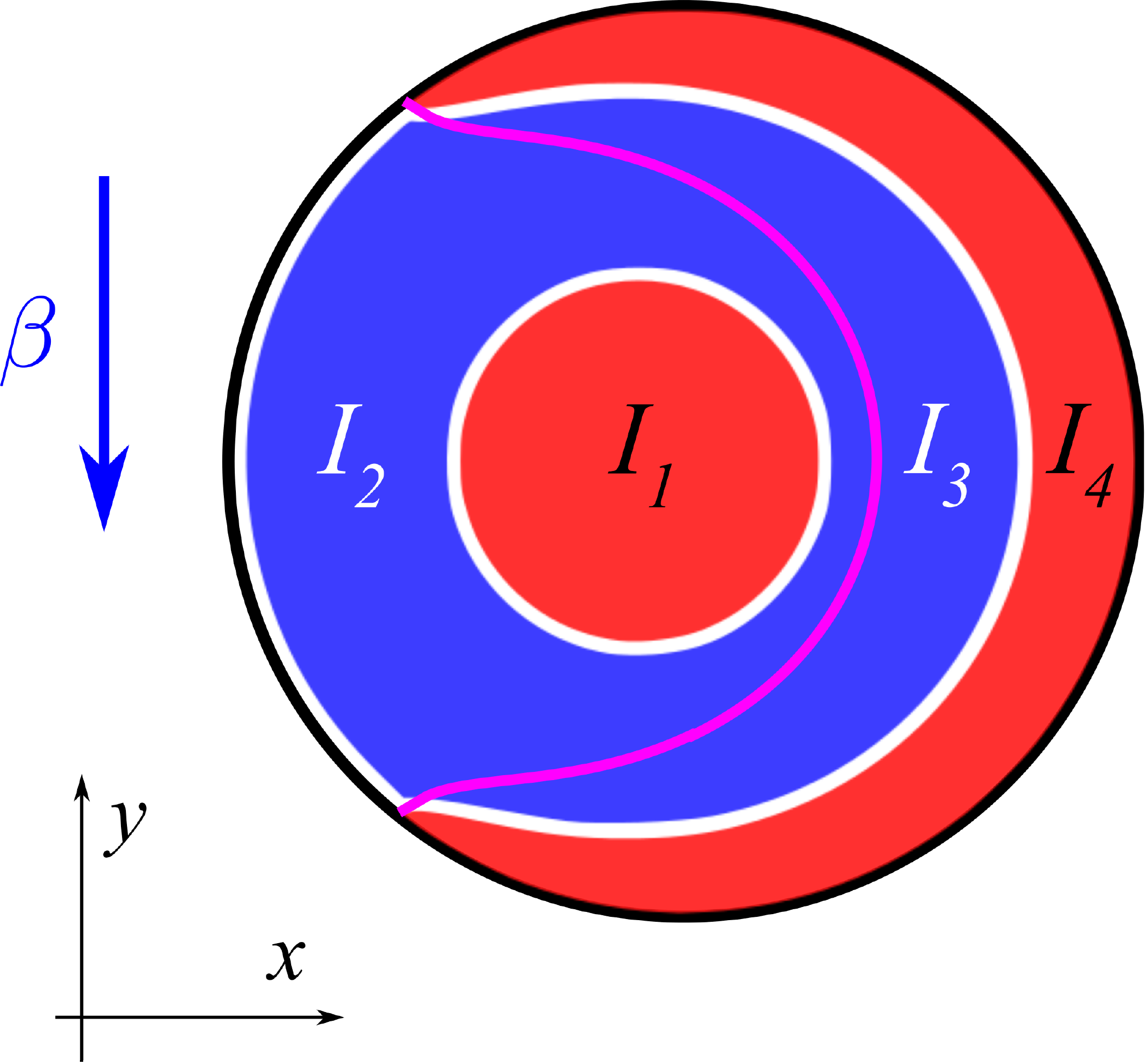}
\end{center}
\caption{Schematic diagram illustrating the different integration domains
  as projected in the $(x,y)$ plane and needed to compute
  expression~\eqref{int}. We recall that the domains reported in the
  figure refer to a rotating black hole with spin in the positive $z$
  direction and moving in the negative $y$-direction.}
\label{fig:diagram}
\end{figure}

Let us now go back to the calculation of the luminosity by taking the
integral of the current density $c\rho_{\rm ind}$ at the horizon 
[see also Eq. (25) of~\cite{Lyutikov:2011}]
\begin{equation}
\label{I} I=(2M)^2\int_{\phi_1}^{\phi_2}
d\phi\int_{\theta_1}^{\theta_2} \rho_{\rm ind} \sin\theta d\theta\,,
\end{equation}
to find the current that flows within each domain restricted by the
surfaces with $E_{\parallel}=0$ and $\rho_{\rm ind}=0$, with $\phi_{1,2}$
and $\theta_{1,2}$ describing the boundaries of these domains. A
schematic diagram illustrating the different integration domains as
projected in the $(x,y)$ plane is shown in Fig. \ref{fig:diagram}, where
the rotating black hole has the spin along the positive $z$ direction and
is moving in the negative $y$-direction.

Interestingly, the integration in the $\theta$-direction can be taken
analytically to obtain the expression
\begin{widetext}

\begin{align}
\label{int}
I=\frac{B_0M}{8\sqrt{2}\pi}
\Bigg \{& 2\beta\int_{0}^{2\pi}d\phi\cos\phi
\left [-14\theta
+\frac{\sin\theta\cos\theta(9-\cos^2\theta)}{1+\cos^2\theta} - 
\frac{18}{\sqrt{2}}\arctan(\sqrt{2}\cot\theta)\right]\Bigg|_{\theta_1}^{\theta_2}
+  \nonumber \\
&  \frac{a}{M}
\int_{0}^{2\pi}d\phi\left[\frac{\cos\theta(5\cos^4\theta+
\cos^2\theta+16)}{1+\cos^2\theta}
-14\arctan\cos\theta\right]\Bigg|_{\theta_1}^{\theta_2}\Bigg\}\,.
\end{align}

\end{widetext}
This equation extends Eq. (25) of Ref.~\cite{Lyutikov:2011} to the case
of a black hole that is also spinning. It should be noted, however, that in
Eq. (25) of~\cite{Lyutikov:2011}, the integration is performed only in
the quadrant restricted by the condition $E_{\parallel}=0$ and that the
separatrixes for the induced charge density $\rho_{\rm ind}$ are
neglected. Equation (\ref{int}), on the other hand, takes into account
both signs of the parallel electric field $E_{\parallel}$ and the induced
charge density $\rho_{\rm ind}$.

The integration of (\ref{int}) in the $\phi$-direction can only be
achieved numerically to obtain the values of the four currents, $I_1$,
$I_2$, $I_3$ and $I_4$ flowing in four domains of the upper hemisphere
(\cf Fig.~\ref{fig:diagram}). Adopting now the physical description
proposed in the membrane paradigm \cite{Price86a}, the surface
resistivity of a black-hole horizon is $\sim 4\pi$, so that we can
estimate that each of the four currents $I_j$ with $j=1,\ldots 4$, will
be responsible for electromagnetic energy losses of the order of $4\pi
I_j^2$. Summing these contributions we readily obtain the electromagnetic
losses as a function of the spin parameter and black-hole velocity, \ie
\begin{equation}
\label{eq:L} 
L_{_{\rm EM}} = L_{_{\rm EM}} (a; \beta) = 
8\pi \sum_{j=1}^{4}|I_j|^2 \,,
\end{equation}
where additional factor $2$ is the result of the symmetry between upper
and lower hemispheres. 

To fix ideas and after considering a reference magnetic field
$B=10^{4}\,{\rm G}$, we report below the estimates for the
electromagnetic losses relative to some representative cases, \eg for a
non-spinning but moving black hole
\begin{equation}
\label{eq:est_1}
L_{_{\rm EM}} (0; \beta)
=7.2\times10^{44}\,\beta^2
\left(\frac{B}{10^4\,\mathrm{G}}\right)^2 \!\!
\left(\frac{M}{10^8\, M_{\odot}}\right)^2
\!\!\mathrm{erg\ s^{-1}}\,, \\
\end{equation}
or for a stationary but rotating black hole
\begin{equation}
\label{eq:est_2}
L_{_{\rm EM}} (a; 0)
=2.4\times10^{44} \, \frac{a^2}{M^2} 
\left(\frac{M}{10^8\, M_{\odot}}\right)^2 \!\!
\left(\frac{B}{10^4\, \mathrm{G}}\right)^2
\!\!\mathrm{erg\ s^{-1}}\,, \\
\end{equation}
or for the more generic case of a black hole with $a=M/2$ and $\beta=0.4$
\begin{equation}
\label{eq:est_3}
L_{_{\rm EM}} (0.5;\,0.4) = 1.9\times10^{44}\,
\left(\frac{B}{10^4\, \mathrm{G}}\right)^2 \!\!
\left(\frac{M}{10^8\, M_{\odot}}\right)^2
\!\!\mathrm{erg\ s^{-1}}\,. \\
\end{equation}

\begin{figure}
\includegraphics[width=0.48\textwidth]{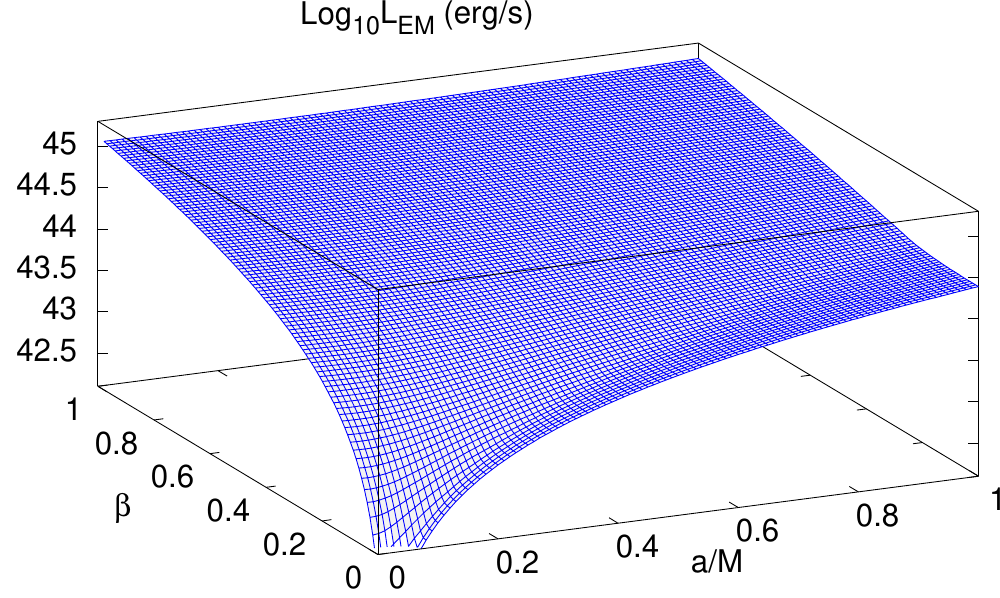}
\caption{Logarithm of the electromagnetic energy losses from the rotating
  and linearly moving black hole immersed into the asymptotically uniform
  magnetic field, estimated as (\ref{eq:L}), for the different values of
  $a$ and $\beta$.}
\label{Fig6}
\end{figure}

By comparing expressions~\eqref{eq:est_1}--\eqref{eq:est_3} it is quite
clear that the order-of-magnitude estimate of the luminosity $\sim
10^{44}\,\mathrm{erg\ s^{-1}}$ is rather robust and does not depend
sensitively on the specific values of $a/M$ and $\beta$. The general
behaviour of the electromagnetic losses as a function of spin and
velocity is shown in Fig.~\ref{Fig6}, which reports the logarithm of the
luminosity (\ref{eq:L}) for different values of $a/M$ and $\beta$;
clearly, because of the perturbative nature of our approach, the
estimates for $a/M \simeq 1$ and $\beta \simeq 1$ should be taken as
indicative only.

It should be noted that for a system with equal values of $\beta$ and $a$
(as it is possible, for instance, in the case of black-hole binaries),
the contribution of the linear velocity will be larger than that of the
angular momentum. This may be explained intuitively if one assumes that
both in the case of a spinning black hole and of a moving one what is
relevant is the relative velocity between the magnetic field lines and
the horizon. Under this assumption, it is easy to realize that even for
maximally rotating black holes, the linear velocity of the horizon is far
smaller than the speed of light and hence the spin-induced contributions
are at most a fraction of those coming from the linear motion. A similar
conclusion, namely that linearly moving black holes lose more energy than
spinning ones, was also reached in Ref.~\cite{Neilsen:2010ax}, where
numerical-relativity calculations were carried out for moving and
spinning black holes in a force-free plasma. Of course, from an
astrophysical point of view it is probably easier to produce black holes
that are close to being maximally spinning than black holes that are
moving near the speed of light.

As a validation of the robustness of the estimates in
expressions~\eqref{eq:est_1}--\eqref{eq:est_3}, we can compare them with
the electromagnetic losses derived in terms of the Poynting flux computed
from numerical simulations of the merger of binary black holes in a force
free plasma~\cite{Palenzuela:2010b,Alic:2012}. Of course, this comparison
should be taken with some care, since expression (\ref{eq:L}) for the
energy losses is obtained using the expression for the induced charge
density which, in turn, is derived from the electrovacuum solutions for
the electromagnetic fields (\ref{BKerr}) and (\ref{EKerr}). The presence
of a plasma magnetosphere, as the one emerging from the numerical
simulations, changes the structure of the electromagnetic fields around
the black hole as the currents generated by the charged particles will
create additional components of the magnetic and electric fields,
possibly screening the component of the electric field parallel to the
magnetic one. Bearing this in mind and assuming that the energy losses
from the binary system of black holes to be a sum of the losses from the
individual black holes, it is interesting to note that the estimate in
\eqref{eq:est_3} is only a factor three larger than the one computed in
Ref.~\cite{Alic:2012} during the inspiral of binary system of black-hole
within a force-free plasma; a similar but slightly worse agreement is obtained
also when comparing with the results in
Ref.~\cite{Palenzuela:2010b}. This agreement is of course reassuring, but
is mostly the result of the fact that our luminosity
estimate~\eqref{eq:est_3} has the right scaling properties with mass and
magnetic field, rather than the same level of accuracy of the numerical
simulations.

\section{Conclusions}
\label{CONC}

We have derived analytic solutions for the structure of test
electromagnetic fields in the vicinity of a rotating black hole that is
also moving at constant speed with respect to an asymptotically uniform
magnetic field aligned with the black-hole spin. In practice, this
represents the extension to a moving black hole of the classical
solution found in 1974 by Wald~\cite{Wald:74bh} for a spinning black hole
in a uniform magnetic field. In order to avoid singularities at the
black-hole horizon we have employed Kerr-Schild coordinates instead of
the Boyer-Lindquist coordinates that have been adopted in recent and
related work~\cite{Lyutikov:2011}.

Overall, the electromagnetic fields found can be seen as the linear
combination of a contribution coming from the translational motion of
black hole relative to the magnetic field, together with a contribution
coming from the black-hole rotation and thus ultimately from the dragging
of inertial frames in the Kerr spacetime. These latter contributions
decay more rapidly with the distance from the black hole than those due
to the black-hole velocity. As a result, at large distances from the
black hole, the structure of the electromagnetic fields, \ie the
component of the electric field parallel to the magnetic field and the
induced charge density necessary to screen this component, are rather
similar to those described in Ref.~\cite{Lyutikov:2011}, where a moving
Schwarzschild black hole was considered.

After adopting a membrane-paradigm description of the electrodynamics of
moving and spinning black holes~\cite{Price86a}, we could estimate the
electromagnetic energy losses by computing the currents that would
develop on the black-hole horizon as a function of its spin and
velocity. Our choice of coordinates was particularly useful for this
task, as none of the electromagnetic fields that we have derived is
singular at the horizon, in contrast with what would happen if more
standard Boyer-Lindquist coordinates were used~\cite{Lyutikov:2011}. Not
surprisingly, we find that the electromagnetic losses depend
quadratically on the black-hole dimensionless spin $a/M$ and on the
velocity velocity $\beta$. However, we find that the energy losses due to
the spin are considerably smaller than those due to the translational
motion; this is in qualitative agreement with the conclusions reached in
Ref.~\cite{Neilsen:2010ax}, who carried out numerical-relativity
calculations of moving and spinning black holes in a force-free plasma.

All things considered, an order-of-magnitude estimate for the
electromagnetic luminosity that would be produced by such an object is
$L_{_{\rm EM}} \sim 10^{44}\ \mathrm{erg\ s^{-1}}$, and this seems to be
rather robust and does not depend sensitively on the specific values of
$a/M$ and $\beta$. Interestingly, this estimate is only a factor three
larger than the one computed during the inspiral of a binary system of
black holes in the force-free approximation~\cite{Palenzuela:2010b,
  Alic:2012}. This agreement is mostly due to the correct scaling of our
estimate with mass and magnetic field, but it highlights that analytic
and more idealized modelling of the electrodynamics of rotating black
holes in magnetic fields can be useful approach to model, at least
qualitatively, much more complex and realistic scenarios.

\begin{acknowledgments}
It is a pleasure to thank Maxim Lyutikov for interesting
discussions. This work has been partially supported by the ``CompStar'',
a Research Networking Programme of the European Science
Foundation. Partial support also comes from the DFG grant SFB/Transregio
7 and the Volkswagen Stiftung (Grant 86 866).
\end{acknowledgments}



\appendix
\section{Maxwell equations for a black hole moving in a magnetic field}
\label{Ap1}

In the following we sketch the procedure we have followed to derive the
solutions of the Maxwell equations for a black hole moving in a magnetic
field. We start by considering a spherical polar coordinate system
$(t,r,\theta,\phi)$ in which we rewrite the Schwarzschild metric as
\begin{equation}
\label{metSch} ds^2=-N^2 dt^2+N^{-2}dr^2+r^2(d\theta^2+\sin^2\theta
d\phi^2)\,.
\end{equation}
In such a background spacetime, the vacuum Maxwell equations
(\ref{Max_1})--(\ref{Max_2}) become
\begin{eqnarray}
\label{Pot1} 
&& \hskip -0.5cm
N^2\frac{\partial}{\partial
r}\left(r^2\frac{\partial}{\partial r}A_{t}\right)
+\frac{1}{\sin\theta'}\frac{\partial}{\partial
\theta'}\left(\sin\theta' \frac{\partial}{\partial
\theta'}A_{t}\right)=0\,, ~~ \\  
\label{Pot2} 
&& \hskip -0.5cm
r^2 \frac{\partial}{\partial r}\left(N^2\frac{\partial}{\partial
r}A_{\phi}\right) 
+\sin\theta \frac{\partial}{\partial
\theta}\left(\frac{1}{\sin\theta}\frac{\partial}{\partial
\theta}A_{\phi}\right)=0\,,
\end{eqnarray}
where $\theta'$ and $\theta$ are angles with respect to the axes aligned
with the electric and magnetic fields at infinity (see
\cite{Lyutikov:2011} for a discussion). Following ~\cite{Beskin:2009}, we
look for the solution of the Eq. (\ref{Pot2}) in the form
\begin{equation}
A_{\phi}=\sum_{m=0}^{\infty}g_m(r)Q_m(\theta)\,,
\end{equation}
where $Q_m(\theta)$ are the eigenfunctions of the operator
\begin{equation}
\hat{\mathcal{L}}_{\theta}=\sin\theta \frac{\partial}{\partial
\theta}\left(\frac{1}{\sin\theta}\frac{\partial}{\partial
\theta}\right)\,,
\end{equation}
and the equation for the radial eigenfunctions $g_m(r)$ has the form
\begin{equation}
\label{g} x(x-1)\frac{d^2g_m}{dx^2}+\frac{dg_m}{dx}-m(m+1)g_m=0\,,
\end{equation}
where $x \equiv r/2M$. As argued in Ref.~\cite{Beskin:2009}, the physical
solution of the equation (\ref{Pot2}) corresponding to a magnetic field
$B_0$ that is uniform at the infinity is given by $g_m(x)=x^2
\mathcal{F}(1-m,\ m+2;\ 3;\ x)$, where $\mathcal{F}(a,\ b;\ c;\ x)$ is
the hypergeometric function of the argument $x$ with the generic
parameters $a,\ b,\ c$. In the case of $m=1$, the $\phi$-component of the
vector potential has the simple expressions
\begin{equation}
\label{Aphi} A_{\phi}=\frac{B_0}{2}r^2\sin^2\theta\,, \quad
A^{\phi}=\frac{B_0}{2}\,.
\end{equation}

On the other hand, the solution to Eq. (\ref{Pot1}) can be found assuming
the \textit{ansatz}
\begin{equation}
A_{t}=h(r)f(\theta)\,,
\end{equation}
where the functions $h(r)$ and $f(\theta)$ are given by the equations
\begin{eqnarray}
\left(1-\frac{2M}{r}\right)\frac{1}{h}\frac{d}{dr}
\left(r^2\frac{dh}{dr}\right) &=& K\,,
\nonumber \\
\frac{1}{\sin\theta}\frac{1}{f}\frac{d}{d\theta}\left(\sin\theta\frac{d
f}{d\theta}\right) &= &-K\,,
\end{eqnarray}
with $K$ an integration constant. A physical solution of Eq.
(\ref{Pot1}) describing an electric field $E_0$ that is uniform at
infinity corresponds to the case $K=2$, $h(r)=C_1(r-2M)$ and
$f(\theta)=P_1(\cos\theta')=\cos\theta'$, where $P_n(\theta)$ is the
Legendre polynomial of order $n$. Adjusting the constant $C_1$ to obtain
the value of the electric field at infinity to be $E_0$, it is possible
to obtain the expression for the $t$-component of the vector potential as
\begin{equation}
\label{Athprime} A_{t}=E_0rN^2\cos\theta'\,, \quad A^t=-E_0 r
\cos\theta'\,.
\end{equation}

These considerations have been made for a set of spherical polar
coordinates to the which, for instance, the Boyer-Lindquist coordinates
tend to at large distances. To obtain instead the components of the
vector potential $A^{\mu}$ in Kerr-Schild coordinates it is necessary to
use the transformation matrix between the two coordinate
systems. Calculating the components of the transformation matrix is
tedious but straightforward, especially if one bears in mind that that
the coordinates $r$ and $\theta$ are the same in the Boyer-Lindquist
(\textsf{BL}) and Kerr-Schild coordinate systems (\textsf{KS}), \ie
\begin{equation}
\label{eq:eqcoord}
r[\sf{KS}]=r[\sf{BL}]\,, \qquad \qquad
\theta[\sf{KS}]=\theta[\sf{BL}]\,,
\end{equation}
so that the corresponding diagonal components are equal to one. Indeed,
all of the diagonal components of the transformation matrix are equal to
one. On the other hand, the only nonzero off-diagonal components are
(see, \eg ~\cite{Komissarov2004b})
\begin{equation}
\frac{\partial t [\sf{KS}]}{\partial r [\sf{BL}]}=
\frac{2Mr}{\Delta}\,, \qquad  \quad 
\frac{\partial \phi [\sf{KS}]}{\partial r [\sf{BL}]}=
\frac{a}{\Delta}\,.
\end{equation}

Because of the identities~\eqref{eq:eqcoord}, the components of the
vector potential $A^t$ and $A^{\phi}$ from equations (\ref{Aphi}) and
(\ref{Athprime}) in Boyer-Lindquist coordinates are the same also in
Kerr-Schild coordinates. On the other hand, the components $A_{\mu}$ in
the Kerr-Schild coordinates may be obtained from them lowering indexes
with the help of the metric (\ref{metric}) when $a=0$. 

When the black hole is moving in the negative $y$-direction and
orthogonally to the uniform magnetic field in the $z$-direction, a
comoving observer will measure an electric field which is uniform at
infinity and directed along the $x$-axis. The angle $\theta'$ in
Eqs.~(\ref{Pot1}) and (\ref{Athprime}) represents the polar angle with
respect to the asymptotic electric field (\ie the $x$-axis in our
case). To transform the expressions to the polar spherical coordinate
system used in the rest of the paper and in which the polar axis is along
the asymptotic magnetic field, one needs to apply the transformation
$\cos\theta' = \sin\theta \cos\phi$, resulting in Eq. (\ref{aLut}) (see
also the discussion in Ref.~\cite{Lyutikov:2011}).

\section{Dual transformation of the vacuum Maxwell equations}
\label{ApD}

Following~\cite{Lyutikov:2011}, it is easy to show that electric and
magnetic fields given respectively by expressions
(\ref{Ehatr})--(\ref{Ehatphi}) and (\ref{Bhatr})--(\ref{Bhatphi}) have
the identical structure. To demonstrate this, we start from
(\ref{Ehatr})--(\ref{Ehatphi}) and find the corresponding components of
the electric field in the Cartesian system of coordinates
\begin{equation}
\begin{bmatrix}
     E_{\hat{x}}    \\
     E_{\hat{y}}    \\
     E_{\hat{z}}       
\end{bmatrix} = 
\begin{bmatrix}
     \sin\theta\cos\phi & \cos\theta\cos\phi & -\sin\phi   \\
     \sin\theta\sin\phi & \cos\theta\sin\phi & \cos\phi  \\
     \cos\theta & -\sin\theta & 0    
\end{bmatrix}
\begin{bmatrix}
     E_{\hat{r}}    \\
     E_{\hat{\theta}}    \\
     E_{\hat{\phi}}      
\end{bmatrix} \,.
\end{equation}
Now we can introduce another Cartesian system of coordinates $(x',y',z')$
such that $-\hat{x} \rightarrow \hat{z}'$, $\hat{z}\rightarrow\hat{x}'$,
$\hat{y}\rightarrow\hat{y}'$, and associate with it a spherical system of
coordinates $(r', \theta',\phi')$ with $r' = r$. It follows that the
following relations hold between the angles
\begin{eqnarray}
\cos\theta' &=& -\sin\theta\cos\phi\,, \nonumber \\
\sin\theta'\cos\phi' &=& \cos\theta\,, \nonumber \\
\sin\theta'\sin\phi' &=& \sin\theta\sin\phi\,, \nonumber
\end{eqnarray}
while the components of the electric field satisfy
\begin{eqnarray}
E_{\hat{x}'} &=& E_{\hat{z}}\,, \nonumber \\
E_{\hat{y}'} &=& E_{\hat{y}}\,, \\
E_{\hat{z}'} &=& - E_{\hat{x}}\,. \nonumber
\end{eqnarray}
Performing now the transformation to the ``primed'' spherical 
coordinates
\begin{eqnarray}
\begin{bmatrix}
     E_{\hat{r}'}    \\
     E_{\hat{\theta}'}    \\
     E_{\hat{\phi}'}       
\end{bmatrix} &=& 
\begin{bmatrix}
     \sin\theta'\cos\phi' & \sin\theta'\cos\phi' & \cos\theta'   \\
     \cos\theta'\cos\phi' & \cos\theta'\sin\phi' & -\sin\theta'  \\
     -\sin\phi' & \cos\phi' & 0    
\end{bmatrix}
\begin{bmatrix}
     E_{\hat{x}'}    \\
     E_{\hat{y}'}    \\
     E_{\hat{z}'}       
\end{bmatrix} \,, \nonumber \\
\end{eqnarray}
we obtain
\begin{eqnarray}
E_{\hat{r}'}     &=& -E_0\cos\theta'    \\
E_{\hat{\theta}'} &=& {\tilde{N}}^{-1}\left[E_0\sin\theta' - ({2M}/{r})B_0\sin\phi'\right]   \\
E_{\hat{\phi}'}   &=& -({2M}/{r\tilde{N}})B_0\cos\theta'\cos\phi' 
\end{eqnarray}
Comparing these components with expressions
(\ref{Bhatr})--(\ref{Bhatphi}) for the magnetic field, it is clear that
the electric field has the same structure with respect to the $x$-axis as
the magnetic field with respect to the $z$-axis. Thus, as remarked in the
footnote \ref{footnote:dual}, the dual solution to
(\ref{Ehatr})--(\ref{Bhatphi}) describes a Schwarzschild black hole
moving in the $-y$ direction in an asymptotically uniform magnetic field
directed along the $-x$ direction.

\section{Expression for induced charge density}
\label{Ap2}

We next provide the full expression for the induced charge density
$\rho_{\rm ind}$ calculated using the expressions (\ref{Ehatr}) and
(\ref{Bhatr}) for the electromagnetic field components. In the case of a
non-spinning black hole, the resulting expression at all orders of
$\beta$ is given by

\begin{widetext}
\begin{eqnarray}
\label{rho_full}
\rho_{\rm ind}=\frac{1}{2\pi r^6\tilde{N}}\frac{E_0M\cos\phi\sin\theta
(\mathcal{A}r^4+\mathcal{B}Mr^3+2\mathcal{C}M^2r^2+4\mathcal{D}M^3r+
16\mathcal{E}M^4)}{\left[\tilde{N}-{2M}\sin\theta
(\sin\theta+\beta\sin\phi)/{r}
+{4M^2}\beta^2(1-\sin^2\theta\cos^2\phi)/{r^2}\right]^2}\
,
\end{eqnarray}
where the different coefficients $\mathcal{A}, \mathcal{B}, \ldots
\mathcal{E}$ are shorthands for more extended expressions, \ie
\begin{eqnarray}
&& \mathcal{A} \equiv 3\cos^2\theta-1\,, \nonumber \\
&& \mathcal{B} \equiv \beta
\sin\theta\sin\phi(3-\cos3\theta)+2\cos^2\theta(3+\cos^2\theta) -\sin^2\theta
(1-\cos2\theta)\,,
\nonumber \\
&& \mathcal{C} \equiv \beta^2\left[-\cos^4\theta(1+\cos2\phi)
-2\cos^2\theta\sin^2\phi-2\sin^2\theta(1-\cos2\phi)\right]
\nonumber \\
&& \qquad\qquad
+\beta\sin\theta\sin\phi(1-3\cos2\theta)+4\cos^4\theta+\sin^2\theta(1+\cos2\theta)
,
\nonumber \\
&& \mathcal{D} \equiv 2\beta^3\sin\theta\sin\phi(1-\sin^2\theta\cos^2\phi)+2\beta^2\sin^2\theta(\cos2\phi-1)
-\beta\sin\phi\sin\theta(1+\cos2\theta)\,,
\nonumber \\
&& \mathcal{E} \equiv \beta^3\sin\phi\sin\theta(1-\sin^2\theta\cos^2\phi)\,.
\end{eqnarray}

\end{widetext}

\section{Spacetime metric in slow rotation approximation}
\label{Ap3}

Finally, we provide explicit expressions for the metric tensor and the
components of the orthonormal ZAMO tetrad $\boldsymbol{e}^{\hat{\mu}}$ in
the slow-rotation approximation. More specifically, at first order in the
dimensionless angular momentum $a/M$, the metric tensor components
(\ref{metric}) take the form
\begin{eqnarray}
\label{metric_lin} && g_{tt}=-N^2\,, \qquad \qquad
g_{tr}=g_{rt}=\frac{2M}{r}\,, \\&& g_{t \phi}=g_{\phi
t}=-\frac{2Ma\sin^2\theta}{r}\,, \quad g_{rr}=1+\frac{2M}{r}\,, \quad
g_{\theta\theta}=r^2\,, \nonumber \\ && g_{r \phi}=
g_{\phi r}=-a\tilde{N}^{2}\sin^2\theta\,, \quad
g_{\phi\phi}=r^2\sin^2\theta\,, \nonumber
\end{eqnarray}
while the tetrad has contravariant and covariant components given
respectively by
\begin{widetext}
\begin{eqnarray}
\label{tetrads1}
\left(\begin{array}{llll} e_{t}^{\hat{t}} & e_{t}^{\hat{r}} & 
e_{t}^{\hat{\theta}} & e_{t}^{\hat{\phi}} \\ \\
e_{r}^{\hat{t}} & e_{r}^{\hat{r}} & e_{r}^{\hat{\theta}} &
e_{r}^{\hat{\phi}} \\ \\
e_{\theta}^{\hat{t}} & e_{\theta}^{\hat{r}} & e_{\theta}^{\hat{\theta}} &
e_{\theta}^{\hat{\phi}}\\ \\
e_{\phi}^{\hat{t}} & e_{\phi}^{\hat{r}} & e_{\phi}^{\hat{\theta}} &
e_{\phi}^{\hat{\phi}}\\ \\
 \end{array}\right)&=& 
\left(\begin{array}{cccc}
\dfrac{1}{\tilde{N}} & \dfrac{2M}{r\tilde{N}} &
0 & -\dfrac{2Ma\sin\theta}{r^2} \\ \\
0 & \tilde{N} & 0 & -\dfrac{a\tilde{N}^{2}\sin\theta}{r} \\ \\
0 & 0 & r & 0\\ \\
0 & 0 & 0 & r\sin\theta \\ \\
 \end{array}\right)\,, 
\qquad \qquad
\left(\begin{array}{llll} 
e^{t}_{\hat{t}} & e^{t}_{\hat{r}} & e^{t}_{\hat{\theta}} &
e^{t}_{\hat{\phi}} \\ \\
e^{r}_{\hat{t}} & e^{r}_{\hat{r}} & e^{r}_{\hat{\theta}} &
e^{r}_{\hat{\phi}} \\ \\
e^{\theta}_{\hat{t}} & e^{\theta}_{\hat{r}} & e^{\theta}_{\hat{\theta}} &
e^{\theta}_{\hat{\phi}} \\ \\
e^{\phi}_{\hat{t}} & e^{\phi}_{\hat{r}} & e^{\phi}_{\hat{\theta}} &
e^{\phi}_{\hat{\phi}}\\ \\
 \end{array}\right) = 
\left(\begin{array}{cccc}\tilde{N} & 0 & 0 & 0 \\ \\
-\dfrac{2M}{r\tilde{N}} & \dfrac{1}{\tilde{N}} & 0 & 0 \\ \\
0 & 0 & \dfrac{1}{r} & 0\\ \\
0 & \dfrac{a \tilde{N}}{r^2} & 0 & \dfrac{1}{r\sin\theta}\\ \\
 \end{array}\right)\,. \nonumber \\
\end{eqnarray}
\end{widetext}

\bibliographystyle{apsrev4-1-noeprint}
\bibliography{aeireferences}

\end{document}